\documentclass[12pt]{article}

\usepackage[utf8]{inputenc}
\usepackage{amsmath}
\usepackage{amsthm}
\usepackage{amsfonts}
\usepackage{mathtools}
\usepackage{amssymb}
\usepackage{comment}
\usepackage{algorithm}
\usepackage{algpseudocode}

\usepackage{dsfont}
\usepackage{stackengine}
 \usepackage{booktabs}
 \usepackage{enumitem}
  \usepackage{pdfpages}

\usepackage[hidelinks]{hyperref}

\usepackage{tikz}
\usepackage{dblfloatfix} 
\usepackage[labelfont=bf]{caption}
\usepackage{subcaption}
\usepackage{tkz-tab}
\usepackage{dsfont}
\usepackage{wrapfig}
\usepackage{wasysym}
\usepackage{enumitem}
\usepackage{adjustbox}
\usepackage{ragged2e}
\usepackage{longtable}
\usepackage{changepage}
\usepackage{setspace}
\usepackage{hhline}
\usepackage{multicol}
\usepackage{tabto}
\usepackage{float}
\usepackage{multirow}
\usepackage{makecell}
\usepackage{fancyhdr}
\usepackage{mathdots}
\usepackage{graphicx}

\usepackage{times}

\usepackage{algorithm}
\usepackage{algpseudocode}%

\newenvironment{sciabstract}{%
\begin{quote} \bf}
{\end{quote}}

\usepackage{dsfont}
\usepackage{amsfonts}
\usepackage{amsthm}
\usepackage{amsmath}
\usepackage{amscd}
\usepackage{amssymb}
\usepackage{enumitem}

\usepackage{float}

\usepackage{caption}
\usepackage{subcaption}

\usepackage{comment}

\theoremstyle{definition}

\newtheorem{?}[Th]{Problem}

\usepackage{mathtools}
\mathtoolsset{showonlyrefs}

\usepackage{wasysym}
\usepackage{yhmath}

\title{\LARGE{Extended Persistent Homology Distinguishes Simple and Complex Contagions with High Accuracy}}

\author{Vahid Shamsaddini$^{a}$ and M. Amin Rahimian$^{b,\ast}$ \\
\normalsize{$^a$~Department of Mathematics, Sharif University of Technology \vspace{-1mm}}\\
\normalsize{$^b$~Department of Industrial Engineering, University of Pittsburgh \vspace{-1mm}}\\
\normalsize{$^\ast$~To whom correspondence should be addressed; email: rahimian@pitt.edu.}
}

\date{}

\begin{document}

\maketitle

\vspace{-10mm}

\begin{sciabstract}
The social contagion literature models the spread of ideas, norms, and behaviors across different social networks as simple or complex contagions. Simple contagions are modeled by independent cascade or bond percolation processes that allow every infected node to infect its neighbors at a specific rate. However, complex contagions are modeled using bootstrap or threshold percolation where a single infected neighbor is not enough and the aggregate effect of the infection in a neighborhood can reinforce the contagion beyond the sum of individual neighboring infections. However, distinguishing simple and complex contagions using observational data poses a significant challenge in practice. Estimating population-level activation functions from observed contagion dynamics is hindered by confounding factors that influence adoptions (other than neighborhood interactions), as well as heterogeneity in individual behaviors and modeling variations that make it difficult to design appropriate null models for inferring contagion types. Here, we show that a new tool from topological data analysis (TDA), called extended persistent homology (EPH), when applied to contagion processes over networks, can effectively detect simple and complex contagion processes, as well as predict their parameters. We train classification and regression models using EPH-based topological summaries computed on simulated simple and complex contagion dynamics on three real-world network datasets and obtain high predictive performance over a wide range of contagion parameters and under a variety of informational constraints, including uncertainty in model parameters, noise, and partial observability of contagion dynamics. EPH captures the role of cycles of varying lengths in the observed contagion dynamics and offers a useful metric to classify contagion models and predict their parameters. Analyzing geometrical features of network contagion using TDA tools such as EPH can find applications in other network problems such as seeding, vaccination, and quarantine optimization, as well as network inference and reconstruction problems.

\end{sciabstract}

\noindent
\section*{Introduction}
 How do ideas, innovations, and norms get adopted and spread through large social networks? Social interactions facilitate the spread of behaviors, and understanding the role of contact structure is a central question in network theory for the social and behavioral sciences \cite{daley1964epidemics,sprague2017evidence,ugander2012structural,gronow2021policy}. Characterizing propagation patterns in networks has major implications for understanding the role of contact structure and subsequent problems such as optimal targeting of interventions \cite{kim2015social,schoenebeck2022think,airoldi2024induction,costlyseeding,guilbeault2021topological} and network reconstruction \cite{peixoto2019network,Landry_2024}.  

Two model categories have emerged in the study of social contagion processes \cite{centola2007complex,murphy2021deep,cencetti2023distinguishing,st2024nonlinear}: Simple contagions in which individuals can pass infections on to their neighbors independently at some rate or with some probability ($q$) \cite{anderson1991infectious}; and complex contagions that require social reinforcement, where infections in a neighborhood reinforce each other and a threshold number ($\theta$) of neighboring infected nodes can cause an individual to switch with high probability \cite{lehmann2018complex}.

Different theories have emerged about how the network structure should affect the spread of simple and complex contagions, for example suggesting that reduced clustering (e.g., by replacing ``short", triad-closing ties with random ``long" ones) accelerates simple contagions \cite{watts1998collective,newman1999renormalization}, but inhibits complex ones \cite{centola2007complex,montanari2010spread}. Others suggest a more harmonized view that long ties accelerate both simple and complex contagions by allowing infections below the threshold to occur with a small probability --- denoted by a parameter $q$ that goes to zero with increasing network size \cite{eckles2024long}.



Establishing statistical evidence of simple and complex contagions in observational data is difficult in part due to the confounding of homophily with network adoptions that make it difficult to credibly estimate adoption curves \cite{aral2009distinguishing}. Randomization of exposure of individuals to their neighbors can produce reliable measures of adoption probabilities with increasing numbers of adopters \cite{bakshy2012role}; however, even empirical studies that supposedly provide evidence for complex contagion find a substantial probability of adoption with a single adopting neighbor that complicates the distinction of simple and complex contagions \cite{monsted2017evidence,bakshy2012social,leskovec2007dynamics,ugander2012structural}; see also \cite[SI section S2.1]{eckles2024long}. Indeed, perhaps some of the strongest evidence for complex contagion comes indirectly from comparing the spread rate on lattice structures comprised entirely of short tiers and random regular graphs with very few short ties, for example, from the randomized controlled trail in \cite{centola2010spread}.



The topological structure of the network among the infected nodes offers a promising feature to infer contagion processes and, in particular, to detect simple and complex contagion types. Persistent homology (PH) is a popular tool in topological data analysis (TDA) to measure homology groups of various dimensions in point cloud and graph-based datasets \cite{carlsson2009topology}; the homology group represents the set of topological invariants, or informally holes, of different dimensions. A modification of persistent homology, called ``extended persistent homology'' (EPH), allows us to measure the lengths of cycles as they arise during the contagion process. This extension is particularly useful, as it improves the stability and interpretability of persistence diagrams, allowing a more accurate analysis of cycles and their topological structures \cite{edelsbrunner2008persistent, cohen2009extending}. Using the reverse of the infection time step (that is, taking the time step at which a node is infected and multiplying it by $-1$), we construct an EPH filtration to measure the length of cycles that emerge during the contagion process. We test the potential of this EPH-based feature for detecting both simple and complex contagion, as well as for regressing the parameters of contagion processes simulated on empirical networks across a range of contagion settings.

EPH is more suitable for detecting simple and complex contagions than standard PH because when applied directly to network structures, topological cycles (loops) often exhibit infinite persistence, which prevents us from effectively measuring the transient topological features to distinguish different contagion mechanisms. In contrast, EPH offers a framework for capturing the lifespan of such features and provides relevant topological summaries that we can use to characterize contagion processes in terms of these lifespans. Furthermore, we propose to use the graph structure directly as the underlying topological space for constructing EPH-based features, rather than using the graph structure to construct simplicial complexes (such as clique complexes). This decision was based on the rationale that the native graph topology itself may more faithfully represent the constraints and pathways governing the propagation of the contagion, especially when one is concerned solely with dyadic interactions (ignoring hyperedges and higher-order group interactions).

Although the question of how structure affects the spread of contagion occupies a central position in network science \cite{granovetter1973strength,centola2007complex,ghasemiesfeh2013complex,eckles2024long}, the reverse question of whether different contagion processes generate distinct infection patterns remains mostly unexplored. In \cite{contreras2024infection}, Contreras, Cencetti, and Barrat find that the infection patterns in simple contagion remain invariant between parameters. However, in complex contagion models, altering parameters, such as the threshold, significantly affects the patterns of network diffusion. Our regression analysis suggests that the proposed topological feature based on EPH varies substantially with both simple and complex contagion parameters ($q$ and $\theta$), and can be used to predict these parameters from the observed infection patterns.

Compared directly to our work, Cencetti et al. in \cite{cencetti2023distinguishing} use the correlation between the infection order of network nodes with their local topology (degrees of nodes) to distinguish simple and complex contagion processes, as well as other contagion processes involving higher-order group interactions. We use the reversed infection time step of the nodes as filtration to construct an EPH-based feature that outperforms the correlation-based feature of Cencetti et al. \cite{cencetti2023distinguishing} in classifying simple and complex contagion mechanisms.

Also related to our methodology, the authors in \cite{taylor2015topological} use persistent homology (PH) to examine how contagions spread across networks characterized by short and long distance connections. Their approach for constructing contagion maps using PH facilitates a deeper understanding of the underlying manifold structures of networks that can improve the prediction, modeling, and control of contagion dynamics.

Various applications of mathematical topology are being explored for better understanding network contagion. For example, the authors of \cite{guilbeault2021topological} use topological insights to redefine concepts of centrality and distance metrics to better suit complex contagion models. Considering the seeding problem, using this approach, it is possible to identify central nodes as suitable candidates for seeding complex contagions.
The authors in \cite{iacopini2019simplicial} apply simplicial complexes from topology to define a new approach to modeling contagion processes based on group interactions that go beyond pairwise connections between nodes.

An important open problem is how to study the interaction between topological features and contagion dynamics in a network. By this, we mean that while certain topological features, such as cycles or clusters, are known to be important, it is crucial to analyze them when a specific contagion process is happening on the network. Therefore, we modify the filtration of extended persistent homology for a specific contagion model to construct a new EPH-based topological feature that is adapted to the observed contagion dynamics. We use this new topological feature to classify simple and complex contagions and predict their parameters. Our classifier can detect simple and complex contagions with high accuracy, even with limited information, for example, observing only a few contagion steps or a small subset of nodes.

\section*{Results}
We model both simple and complex contagion dynamics using the susceptible-infected (SI) framework with a zero recovery rate. Specifically, we focus on absolute threshold-based contagion, where a node becomes infected when the count of its infected neighbors exceeds a fixed threshold $\theta$ \cite{cencetti2023distinguishing}. Furthermore, following the noisy threshold-based contagion framework of Eckles et al. \cite{eckles2024long}, we allow nodes with at least one infected neighbor to have a (small) probability ($q$) of becoming infected below the threshold ($\theta$). To signify their connection in our results, we denote simple contagion dynamics by $\theta = \infty$. We terminate simulations of contagion processes once $85\%$ of the network is infected. This serves as the default termination criterion for all simulations, and we explicitly state otherwise when testing alternative termination criteria. Additional details on these contagion models are available in the Methods section.

\begin{figure}[H]
    \centering

    \begin{subfigure}[b]{0.18\textwidth}
        \centering
        \includegraphics[width=\textwidth]{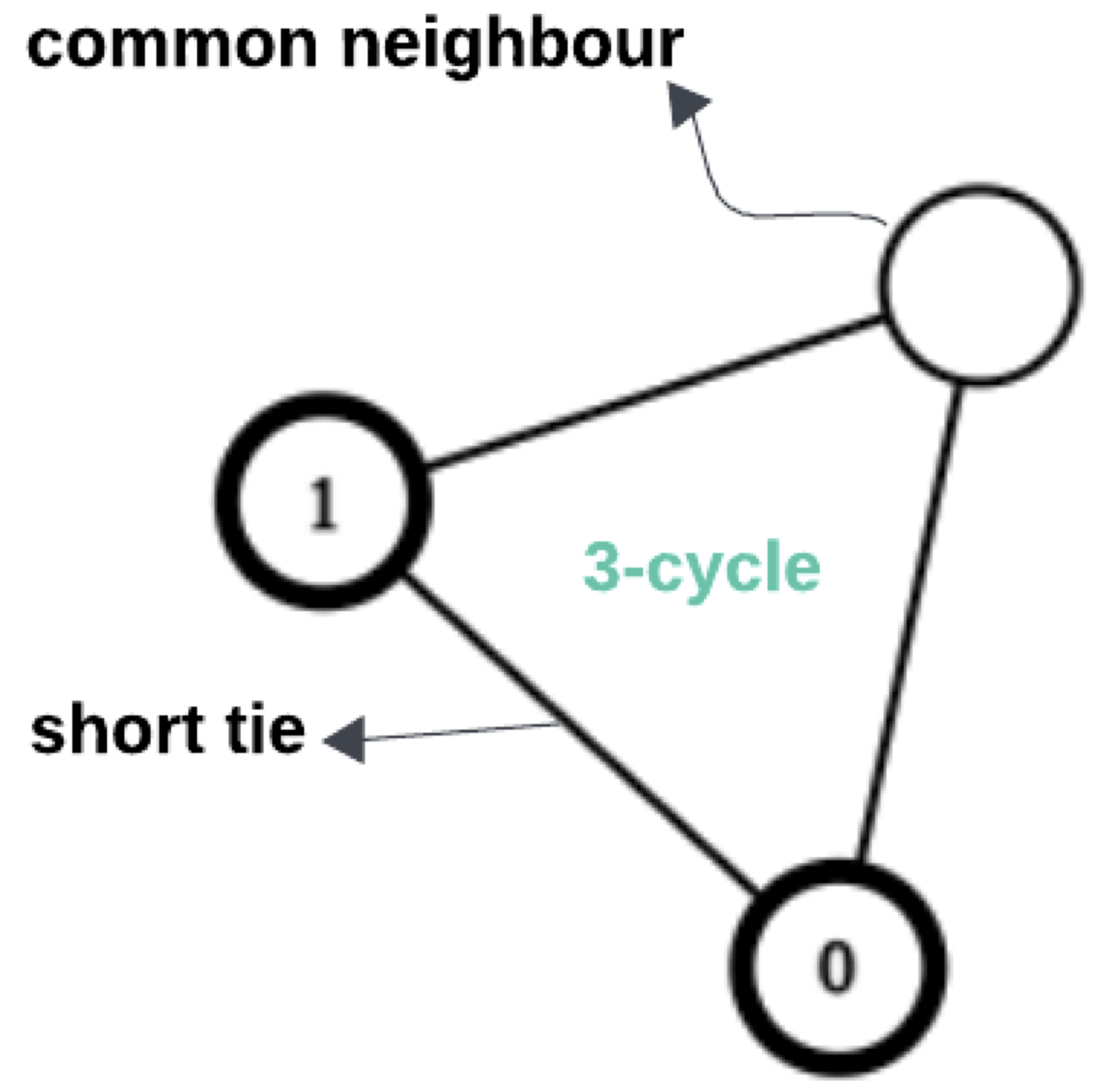}
        \caption{}
        \label{fig:panel1}
    \end{subfigure}
    \hfill
    \begin{subfigure}[b]{0.23\textwidth}
        \centering
        \includegraphics[width=\textwidth]{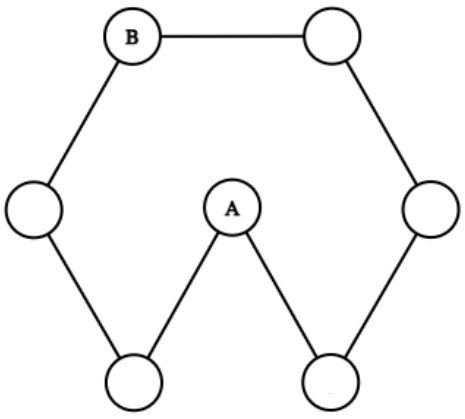}
        \caption{}
        \label{fig:panel3}
    \end{subfigure}
    \hfill
    \begin{subfigure}[b]{0.23\textwidth}
        \centering
        \includegraphics[width=\textwidth]{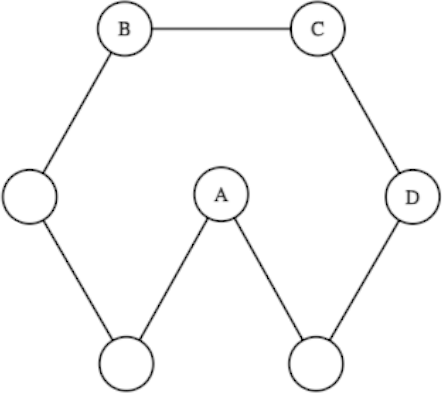}
        \caption{}
        \label{fig:panel4}
    \end{subfigure}
    \hfill
    \begin{subfigure}[b]{0.23\textwidth}
        \centering
        \includegraphics[width=\textwidth]{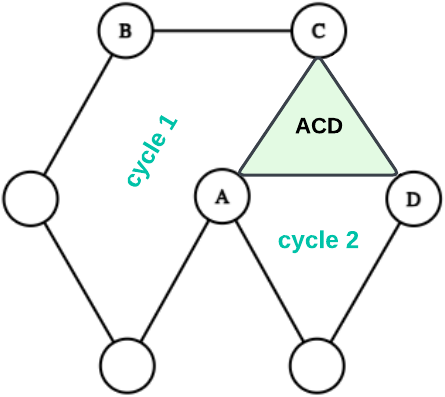}
        \caption{}
        \label{fig:panel5}
    \end{subfigure}
    \caption{\textbf{Panel \ref{fig:panel1}} shows how to determine whether a tie is long or short based on the length of the cycle it creates. A 3-cycle indicates a short tie, and longer cycles signify longer ties (assuming that there are no edges crossing the cycle). \textbf{Panel \ref{fig:panel3}:} To redefine cycle length for network contagions, consider a simple contagion in a cycle with node A initially infected, where the infection spreads with certainty ($\theta = \infty$, $q = 1$). 
    In this example, the time required for all nodes to become infected equals half the cycle's length (path length between A and B). However, to better reflect contagion dynamics, we propose using the total infection time as an extended definition of the cycle length, rather than the traditional cycle length that counts the number of nodes. This extension allows for a more accurate representation of the contagion process in relation to the cycle lengths. In more complex scenarios, such as multiple initially infected nodes, this definition must be adjusted again to account for the varied dynamics of contagion. \textbf{Panel \ref{fig:panel4}:} For example, if nodes A, C and D are all initially infected, then the infection spreads not just from node A, but from the collective influence of nodes A, C, and D together. \textbf{Panel \ref{fig:panel5}:} To accurately model the spread dynamics from the collective influence of nodes A, C, and D, we can treat the initial infected nodes A, C, and D as a single entity. This grouping divides the original cycle into two other independent cycles (labeled cycle 1 and cycle 2 in Figure \ref{fig:panel5}), where the contagion spreads separately. Using the reverse of the infection steps as filtration in EPH results in a scenario that is completely similar to the one shown (considering initially infected nodes as a single block). This demonstrates why we use EPH with the negative infection order as filtration.}
    

\label{fig:intuition_of_method}
\end{figure}

The foundation of our idea lies in the length of cycles (loops) in networks as crucial topological information for studying contagion (see Figure \ref{fig:intuition_of_method}). Why are cycles important? This is related to the concept of long and short ties \cite{granovetter1973strength,centola2007complex,jahani2023long,eckles2024long}. Short ties in static networks occur where nodes share common neighbors, forming short cycles with three nodes (Figure \ref{fig:panel1}). However, we need to extend the concept of cycle length to include information on contagion dynamics. For example, a triangle may still indicate a long-range contagion if it includes nodes that are infected at significantly different times. Therefore, we adjust the definition of the cycle length to reflect information on the dynamics of the contagion within the cycle. To develop and implement this idea using algebraic topology, we employ extended persistent homology (EPH) with the reverse of node infection steps as its filtration. Then we calculate the average of the lifetimes of all generators as our topological feature and call it the EPH feature (Figure \ref{fig:intuition_of_method}). In SI section \ref{secSI:cycle_contagion}, we provide more details on how we measure cycle lengths for network contagions. In SI section \ref{secSI:PHandEPH}, we provide mathematical details on the extended persistent homology and its relationship to cycle length and contagions. 

We used several empirical network datasets to perform simulations and evaluate the performance of our proposed EPH feature to infer contagion dynamics. For the classification of simple and complex contagions, we choose the algorithm from \cite{cencetti2023distinguishing} as the baseline to compare our results. Of all the features described in \cite{cencetti2023distinguishing}, we specifically consider the correlation between the order of infection of the nodes and the degrees of the nodes, as this is the metric they introduced to detect simple and complex contagion. Other features discussed in their study, which address higher-order infections, are not applicable here. 

For selecting datasets, we initially chose two datasets that are used in \cite{cencetti2023distinguishing} for direct comparison. These two datasets encode interactions that have occurred in a conference and school context. Additionally, we incorporated another email dataset with a larger number of nodes and edges but without hyperedges, focusing solely on pairwise connections for our study. We provide additional details about these three datasets in the Methods section. We chose the email dataset for our main results because it has the highest number of nodes compared to the other datasets. We provide our results for the remaining datasets in SI section \ref{SIsec:other-results}.

\begin{figure}[H]
    \centering

    \includegraphics[width=1\textwidth]{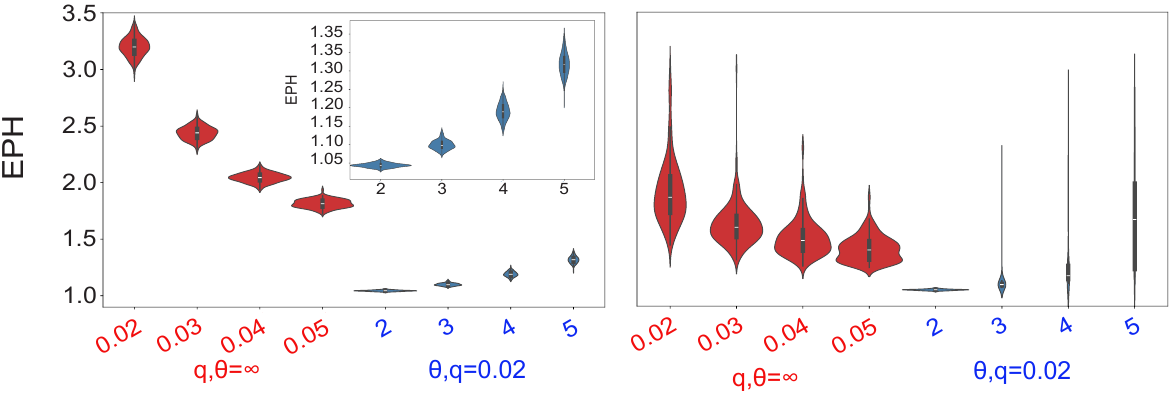}

    \caption{On the left, violin plots showing the distribution of Extended Persistent Homology (EPH) values under simple (red) and complex (blue) contagion models across varying values of threshold $\theta$ and simple contagion probability $q$. Each distribution is derived from 200 simulations on the Email dataset, with the contagion process terminated once $85\%$ of the nodes were infected. For complex contagion ($\theta < \infty$), increasing $\theta$ leads to higher EPH values, reflecting greater temporal dispersion in infection times due to more stringent activation requirements. A similar trend is observed in simple contagion ($\theta = \infty$) as $q$ decreases. On the right, EPH distributions for the same parameter settings, with simulations halted after three infection steps. Compared to simple contagion, complex contagion exhibits greater sensitivity in both the shape and variance of the EPH distribution under parameter changes, highlighting its increased topological and geometric variability.}
    \label{fig:distributions}
\end{figure}

We simulated simple and complex contagions in the email dataset, considering noisy threshold-based complex contagions to be classified against simple contagions. We calculated our topological feature using the extended persistent homology algorithm. Numerical implementation details for computing expending extended persistent homology and simulating simple and complex contagion dynamics are provided in SI section \ref{secSI:details_implementation}. In Figure \ref{fig:distributions}, we plot the distribution of the proposed EPH feature for simple and complex contagions with a range of model parameters. The distinct shapes of these distributions indicate that this feature can effectively differentiate between simple and complex contagions and can also be used in regression analysis to predict the contagion parameters. In particular, the values of the EPH-based topological summaries are generally lower for complex contagion compared to simple contagion. This is consistent with our expectation that shorter cycles, measured through their EPH lifespans, should predominate complex contagion dynamics (Figure \ref{fig:intuition_of_method} and SI section \ref{secSI:cycle_contagion}). Figure \ref{fig:distributions} further indicates that EPH values decrease with increasing value of $q$ for simple contagion and increase with increasing value of $\theta$ for complex contagion. This behavior is also expected, because increasing $q$ leads to faster simple contagions with cycles that have shorter EPH lifespans, while increasing $\theta$ slows down complex contagion and generates cycles with longer EPH lifespans. We obtain similar results for other datasets and report them in the SI section \ref{secSI:distributions}.

As suggested in Figure 2, EPH can serve as a distinguishing feature to classify simple and complex contagions and also to predict the parameters of the contagion. To test this hypothesis, we implemented a classification and a regression framework under the following conditions. We used a decision tree classifier and a polynomial regression function with EPH as their only input feature. The choice of the random forest classifier is consistent with \cite{cencetti2023distinguishing} for our baseline comparisons. In sequel, we present results based on a $80\%-20\%$ train-test split. After training, we evaluated the performance of the classifier in the test set by computing accuracy along with the $95\%$ confidence interval. All contagion simulations began with $1\%$ of the nodes initially infected at random and end after reaching $85\%$ of the nodes. 




\begin{figure}[H]
    \centering
     \includegraphics[width=1\textwidth]{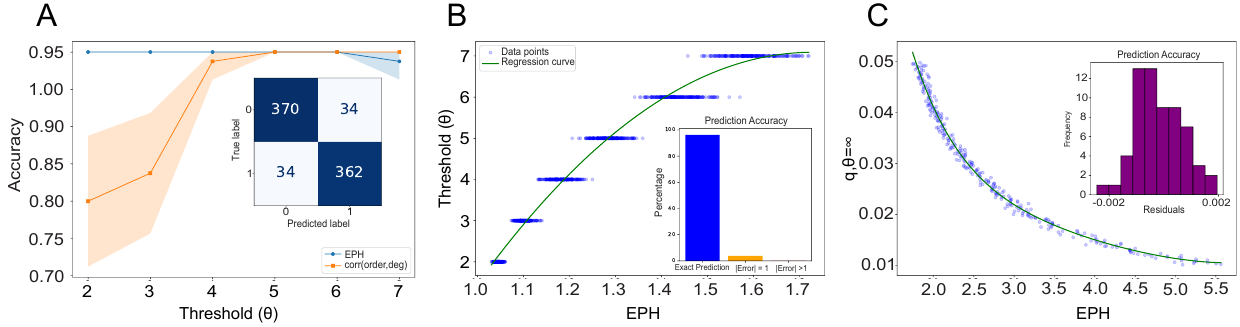}
    
    \caption{\textbf{Panel 3A:} Classification accuracy (with $95\%$ confidence intervals) as a function of the threshold $\theta$ for distinguishing complex contagion types using EPH features. Models were trained and tested on data generated under the same $\theta$ values. Inset: Confusion matrix for a binary classifier differentiating between simple and complex contagion based on EPH. Each contagion type was simulated 800 times. For simple contagion, $q$ was randomly selected from $\{0.02, 0.03, 0.04, 0.05\}$. For complex contagion, $q$ was fixed at $0.02$, and $\theta$ was varied over $\{2, 3, \dots, 7\}$. \textbf{Panel 3B:} Polynomial regression (using second-order polynomials) of threshold $\theta$ on EPH for complex contagion. Inset: Prediction accuracy measured as the percentage of samples for which the regressor predicts $\theta$ exactly, within $\pm 1$, and with absolute error greater than one (no cases were identified in the latter category). \textbf{Panel 3C:} Polynomial regression (using third-order polynomials) of transmission probability $q$ on EPH under simple contagion. Inset: Residual plot showing deviation of predicted values from observed $q$.}
    \label{fig:results}
\end{figure}

As a first step, we compare our results with the baseline in \cite{cencetti2023distinguishing} for the classification problem. We evaluated the robustness of our algorithm in a noisy complex contagion environment where infections below the threshold occur with probability $q$ and analyzed the impact of different threshold values on the performance of our method and the baseline. The results of this analysis are shown in Figure \ref{fig:results} panel A. Here, our approach proves superior in most thresholds. However, at very high values of $\theta\geq 7$ the threshold condition for complex contagion is rarely satisfied during the contagion process; and therefore, complex contagions at high threshold values become very difficult to distinguish from simple contagion based on their EPH-based topological feature. This issue is also apparent in Figure \ref{fig:distributions}, where the distribution of EPH values at high thresholds for complex contagion becomes wider and approaches that of simple contagion. 

In addition to classifying simple and complex contagions, the EPH-based feature can also be used to predict the corresponding contagion parameters: $q$ for simple contagion and $\theta$ for complex contagion. As illustrated in Figure \ref{fig:results} panel B, EPH accurately predicts $\theta$ using polynomial regression. Here again, the ability of the EPH-based topological feature to distinguish different threshold values decreases with increasing thresholds; albeit threshold values can be predicted with perfect precision over a realistic range $(2\leq \theta \leq 5)$. EPH-based regression provides similarly precise predictions for the value of $q$ in simple contagions in panel C of Figure \ref{fig:results}. Here, the shape of the regression curve is reminiscent of $1/q$, which can be explained by the mean of the geometric random variables that represent the wait times for the contagion to spread over the length of a given cycle when the probability of passing infections on each edge is $q$.

Limited information scenarios reflect real-world constraints in data collection and analysis. We evaluated our method under two types of information limitation: having incomplete and inaccurate information about the network structure or having only partial network information about a subset of nodes. In the first scenario, we train the classifier on the merged dataset of two networks and test it on data from a different network, modeling situations where network connections can change and may not be available outside of the contagion context (for example, imagine contact tracing where only connections among infected individuals are revealed and the underlying network structure may change between different disease cascades). In the second scenario, we consider having access only to a random subset of nodes (and their induced subgraph), which we express as an observed ratio (in percentage of nodes). This can be to reduce the cost of data collection by limiting observations to a small fraction of nodes or to reduce computational costs. The latter is important because the time complexity of our method is primarily determined by the persistent homology calculation, which requires reducing the boundary matrix and has a cubic time complexity with respect to the number of edges. We can significantly reduce computational time by considering only a subset of nodes. SI section \ref{secSI:details_implementation:eph} includes additional numerical implementation details of EPH computations.  

\begin{figure}[H]
    \centering

     \includegraphics[width=1\textwidth]{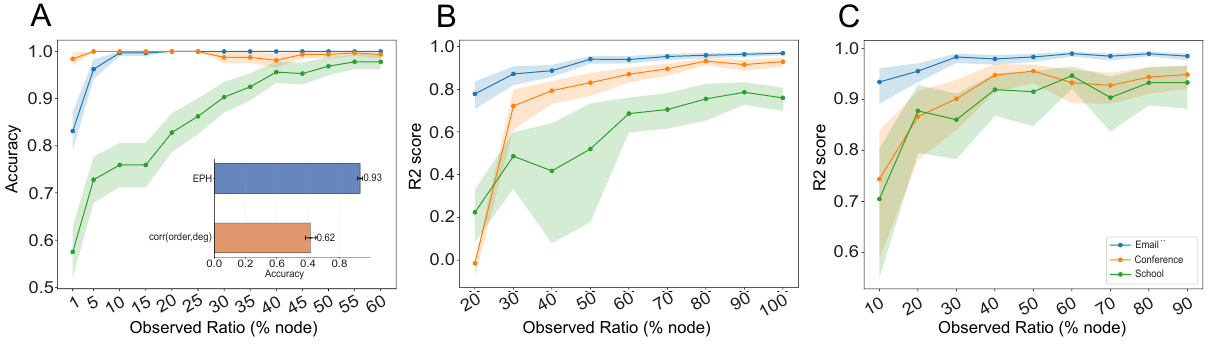}

    \caption{\textbf{Panel 4A:} Classification performance as a function of the percentage of nodes with known infection times. The x-axis indicates the proportion of observed nodes, while the y-axis shows classification accuracy with $95\%$ confidence intervals. Results show that observing approximately $40\%$ of the nodes is sufficient to achieve over $80\%$ classification accuracy for distinguishing between simple and complex contagion using EPH across all datasets. Inset: Generalization performance of our EPH-based classifier compared to a baseline, when training and testing are conducted across three different networks. Our method maintains high accuracy ($\geq 90\%$), while the baseline classifier performs near chance level ($\approx 60\%$). The classifier setup matches that used in previous figures.
\textbf{Panel 4B}: Regression performance for estimating the threshold parameter $\theta$ under complex contagion, as a function of observed node fraction. The y-axis reports the coefficient of determination ($R^2$ score) with $95\%$ confidence intervals.
\textbf{Panel 4C:} Same as Panel B, but for estimating the infection rate $q$ in simple contagion scenarios. For all complex contagion simulations of this figures we set $q=0.02$.
    }
    \label{fig_liminf:results}
\end{figure}

Figure \ref{fig_liminf:results}A demonstrates the robust performance of our method in classifying simple and complex contagions in both of these limited information scenarios. In the first scenario, we only tested classification performance in mixed datasets (Figure \ref{fig_liminf:results}, panel A, inset), because different networks generate different regression curves that cannot be combined to accurately predict contagion parameters in mixed datasets. Figures \ref{fig_liminf:results}B and \ref{fig_liminf:results}C show the robust regression performance to predict the parameters of simple and complex contagion with partial observation of the nodes in three different datasets. For both classification and regression, most of the predictive power is retained with only $60\%$ of the nodes in the observation set. These evaluations assess the robustness of our approach when faced with partial network information or incomplete information about the structure, mirroring challenges encountered in practical applications. 



In summary, we evaluated classification and regression performance using our EPH-based topological feature under various scenarios: noisy complex contagions with infections below the threshold and with only limited information about the network structure and contagion dynamics. The results demonstrate the robust efficacy of our topological feature in capturing the geometry of the contagion dynamics to accurately distinguish simple and complex contagions.

\section*{Discussion}

Both our method and the baseline method use the order of infection of the nodes during the contagion processes to distinguish simple and complex contagions. The baseline method simply uses the correlation of this order with the degrees of the nodes, which represents a basic geometrical aspect of the networks. In this article, we argue for a more sophisticated approach by integrating a geometrical feature of networks with the order of infections. We use extended persistent homology with the reversed order of infections as filtration. This approach captures more comprehensive data from the network compared to simply using node degrees. Our results confirm that this method performs better in detecting simple and complex contagion (Figure \ref{fig:results}, panel A) and can also be used to accurately predict the contagion parameters in each category (panels B and C of Figure \ref{fig:results}). One might have expected that increasing the complex contagion threshold would make it better distinguished from simple contagion (moving away from a minimally complex contagion at $\theta = 2$), but, perhaps unexpectedly, our results indicate that noisy complex contagions are harder to distinguish at higher threshold values because the threshold conditions in those cases are less frequently satisfied, and the observed contagion pattern at high thresholds is mostly shaped by noisy adoptions below threshold, making higher thresholds less distinguishable from each other and from simple contagion dynamics. 

Although this ability to distinguish simple and complex contagion with high accuracy is important in itself, we expect the way that we incorporate geometrical information in measuring contagion dynamics to offer a novel approach to integrating geometry with the contagion processes and a new perspective for analyzing contagion processes on different network structures. This could serve as a starting point for developing new solutions to various contagion-related challenges, such as seeding \cite{costlyseeding,rahimian2023seeding} and network reconstruction \cite{Landry_2024}.



Our analysis opens several avenues for future work. One direction could involve developing a more effective filtration strategy for real scenarios where the exact infection steps of the nodes are unknown, but other information is available. Another important area could involve extending our approach to other types of contagion, such as simplicial contagions and higher-order network interactions. This could necessitate exploring persistent homology of orders higher than one, potentially offering more nuanced insights into the topological structure of contagion processes. Other future research directions could explore the applicability of our method to continuous dynamics, as opposed to the discrete-time dynamics currently studied. This extension could involve various types of contagion with real-valued indices. 

Further research could explore more sophisticated methodologies for extracting and analyzing information derived from extended persistent homology. Although the current study focused on the average lifetime of topological features (generators) as a primary summary statistic, alternative approaches may yield richer insights into contagion dynamics. For example, the application of vectorization techniques such as persistent landscapes or persistence images warrants investigation. These methods transform persistence diagrams into finite-dimensional vector representations suitable for machine learning, potentially capturing more nuanced topological information relevant to contagion processes. Furthermore, exploring a wider array of statistical descriptors derived directly from EPH barcodes (persistence diagrams), beyond the mean lifetime, could offer complementary perspectives and potentially improve classification accuracy or mechanistic understanding.

\section*{Methods} 

\textbf{Models of Contagion.} To model both simple and complex contagion, we use the susceptible-infected (SI) model with a zero recovery rate. In our study of complex contagion processes, we consider both the \textit{absolute} models of threshold contagion. In the \textit{absolute threshold model}, we consider the edges and require that a node is influenced by a fixed number \( \theta \in \{1,2,3,\ldots\} \) of its infected neighbors to be infected. For simple contagion, each node can become infected by contact with a single infected neighbor with a rate $q$ ($\theta = \infty$). 
In addition to absolute threshold models, we also analyze a noisy threshold-based contagion where the contact requirement exhibits stochasticity. In this model, a node becomes infected if at least $\theta$ of its neighbors are infected, matching the deterministic threshold. However, there is also a small probability $q$, that a node will become infected even if only one of its neighbors is infected. This secondary mechanism captures an increase in noise in the contagion process and can potentially make simple and complex contagion harder to distinguish \cite{eckles2024long}.

\noindent\textbf{Extended persistent homology.} Persistent homology (PH) is a technique in topological data analysis that studies data shapes on multiple scales, beginning with the concept of a \textit{filtration}, which is a sequence of spaces \(X_a\), indexed by a parameter \(a\). As \(a\) increases, \(X_a\) expands, revealing new topological features such as loops or voids. These features can appear and disappear; persistent homology tracks these changes and records the lifetime of each feature, summarizing this information in a persistence diagram or barcode. Extended Persistent Homology (EPH) extends this approach by addressing features that persist indefinitely using \textit{relative homology}. It differentiates features that persist as \(a\) continues to increase, capturing those that are integral to the integrity of the structure on all scales, thereby providing a more comprehensive analysis of the data. For more technical details of PH and EPH, please refer to SI section \ref{secSI:PHandEPH}.
\newline

\noindent\textbf{Empirical network datasets.} We have selected three empirical network datasets from three different data collection pipelines:
\begin{enumerate}
    \item Email network: This network is created using email interactions at a major European research institution, representing $1,005$ researchers as nodes with $16,706$ edges reflecting their email exchanges \cite{yin2017local}.
    \item RFID-collected conference data: We analyzed three data sets documenting face-to-face interactions using RFID devices, with a temporal resolution of 20 seconds in three different settings: a workplace, a conference, and a hospital, all detailed in \cite{genois2018can} and \cite{vanhems2013estimating}. We do our analysis on the conference data because it is the largest with $403$ nodes and $9,565$ edges.
    \item Utah school interactions: This dataset, with $591$ nodes and $37,873$ edges, provides a focused view of the original extensive data collection in \cite{toth2015role} that examines interactions between students at a Utah school, captured every 20 seconds using wireless range-enabled nodes.
\end{enumerate}

\section*{Data availability}

The email dataset is publicly available at \cite{yin2017local}. The workplace, conference, and hospital datasets are publicly available at \cite{genois2018can} and \cite{vanhems2013estimating}, and the Utah school dataset is available at \cite{toth2015role}.

\section*{Code availability}
Code for reported simulations can be accessed from \href{https://github.com/shamsvahid2/topology-contagion}{https://github.com/shamsvahid2/topology-contagion}.

\section*{Acknowledgements}{}
We gratefully acknowledge discussions with and feedback from Dean Eckles, Hans Riess and Arnab Sarker. Rahimian
is partially supported by the National Science Foundation under Grants No. 2424684 and No. 2318844. This research was supported in part by the University of Pittsburgh Center for Research Computing, RRID:SCR\_022735, through the resources provided. Specifically, this work used the HTC cluster, which is supported by NIH award number S10OD028483.

\bibliographystyle{Science}

\bibliography{arxive_refrences.bib}

\newpage
\setcounter{page}{1}
\setcounter{equation}{0}
\setcounter{figure}{0}
\renewcommand{\thepage}{S\arabic{page}}
\renewcommand{\thesection}{S\arabic{section}}
\renewcommand{\theequation}{S.\arabic{equation}}
\renewcommand{\figurename}{Supplementary Figure}
\renewcommand{\tablename}{Supplementary Table}
\noindent {\Huge \bf Supplementary Information}\\

\noindent {\small
\tableofcontents}

\newpage

\section{Additional Related Works}\label{secSI:additional-lit-rev} 
Andres et al. (\cite{andres2024distinguishingmechanismssocialcontagion}) investigate the distinguishability of coexisting contagion mechanisms: simple, complex, and spontaneous, using only ego-centric network information. In contrast to previous studies that require complete network knowledge, they formulate the problem as a classification task and apply both likelihood-based inference and random forest classifiers. Through synthetic and empirical experiments, including Twitter data, they demonstrate that local adoption trajectories can be sufficient for mechanism inference. Their results highlight nuanced temporal and structural features that differentiate contagion types, especially with limited observational data. Although we obtain a superior classification accuracy, even with limited observations, our classification problem is significantly different because we treat the contagion type as a property of the network spreading process, not the individual nodes. 

Fink et al. (\cite{Fink_Schmidt_Barash_Kelly_Cameron_Macy_2021}) investigate the observability of complex contagion in empirical Twitter networks by fitting probabilistic adoption models to real-world data. Using 20 popular hashtags from Nigerian Twitter in 2014, they compare the predictive accuracy of simple and complex contagion models under asynchronous user activity. Their results show that complex contagion provides a better fit for politically sensitive hashtags (e.g., AmericaWillKnow), while simple contagion explains most others. Crucially, they demonstrate that traditional threshold-based methods are biased due to temporal sparsity, advocating for model-based inference using adoption curves instead.

Landry et al. (\cite{Landry_2024}) develop a nonparametric Bayesian framework to reconstruct the network structure and contagion dynamics from binary time series, allowing for both simple and complex contagions without assuming a fixed model. Using neighborhood-based SIS dynamics, they show that reconstruction accuracy depends on the interplay between contagion type, network density, and dynamical saturation. In particular, complex contagions outperform simple ones in reconstructing dense or highly clustered networks, whereas simple contagions fare better in sparse settings. Their work highlights that the statistical power of a contagion process to reveal the network structure varies with the complexity and context of the network process.

Sarker et al. (\cite{sarker2024capturingtiestrengthalgebraic}) propose a set of algebraic topological measures derived by Hodge decomposition in simplicial complexes to estimate the strength of the tie in higher-order networks. Their framework captures how structural features such as gradients, curls, and harmonic components relate to local and long-range interactions, effectively reproducing the empirical U-shaped relationship between tie strength and tie range. They further reinterpret Edge PageRank as a stochastic communication process that identifies structurally weak ties that are well-positioned for information diffusion. This work offers a topological reconciliation between Granovetter’s weak-tie theory and recent evidence on strong bridging ties.

Wan et al. (\cite{wan2024randomnessbeatsredundancyinsights}) introduce a stochastic contagion model to examine how probabilistic adoption dynamics influence diffusion in clustered versus random networks. Challenging the canonical view that complex contagions spread more effectively on clustered networks, they show that random networks generally outperform clustered ones across most of the parameter space, even for socially reinforced behaviors. Their analysis reveals that only highly deterministic contagions, those with low baseline and high reinforced adoption probabilities, benefit from clustering. They provide analytical thresholds and simulation results to demarcate the precise conditions under which clustering improves spread, thus refining the theoretical landscape of complex contagion.


\section{Cycle Length and Contagion}\label{secSI:cycle_contagion}

 The primary innovation of our work is to extend the definition of cycle length in graphs within the context of various contagion processes. To understand this extension, we first examine the importance of the cycle length in contagion processes. Subsequently, we elucidate the need to expand the traditional definition of cycle length.

\begin{figure}[H]
\centering
\includegraphics[width=0.24\textwidth]{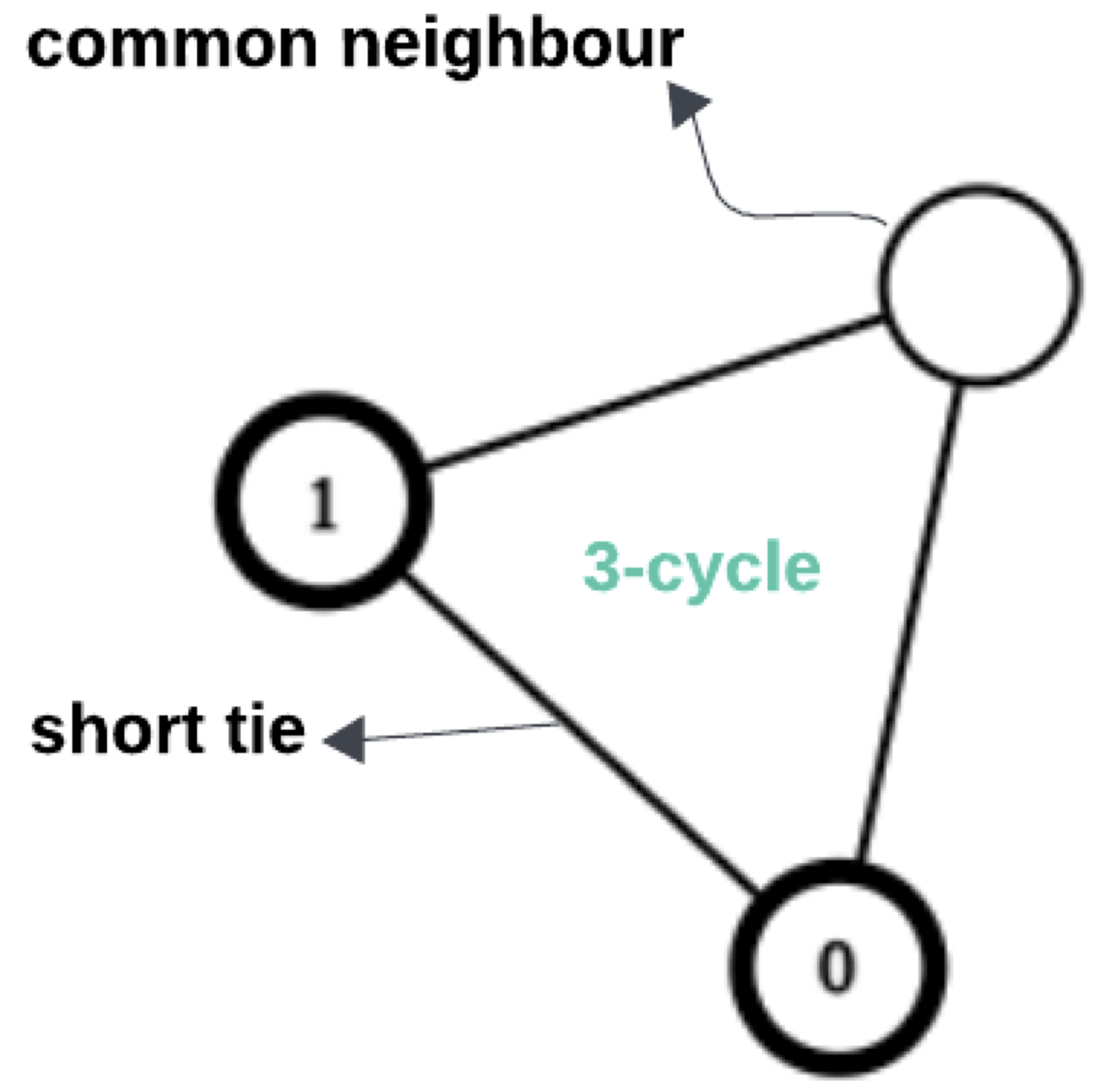}
\caption{\label{short_tie} Short ties indicate 3-cycle}
\end{figure}

 Why are cycles important? This is related to the concept of short and long ties. A short tie occurs where the nodes share a common neighbor, for example, by forming a cycle with three nodes (Supplementary Figure \ref{short_tie}).
Therefore, we can determine whether a tie is long or short on the basis of the length of the cycle it creates. A 3-cycle indicates a short tie, while longer cycles, in cases without additional edges between cycle's nodes, signify longer ties. 

Previous research has explored the relationship between long and short ties in contagion processes. However, a more precise definition is needed to accurately capture the dynamics of contagion. We will explain how we modify the definition of cycle length and propose a new measure of tie length to distinguish long and short ties relevant to contagion dynamics. 


 Consider a contagion process in a network $G = (V,E)$ where $V$ represents the nodes and $E$ the edges. We first define a function $T$ on the nodes, mapping each node to its infection step if it becomes infected or to the last step of infection plus one if it does not. Specifically, if $v \in V$ is infected at step $n$ then $T(v) := n$. For any node $u$ that does not become infected, we set $$T(u) := \max \{T(v)| v \text{ got infected}\} + 1.$$ 
Now, for each cycle, we introduce a new function $L$ that measures the ``contagion" length of $C$ as the difference between the maximum and minimum of the function $T$ on its nodes. Formally, for each cycle $C \subset G$, we set: $$ L(C) = \max \{T(v)| v \in C\} - \min \{T(v)| v \in C\}. $$

\begin{figure}[H]
\centering
\includegraphics[width=0.25\textwidth]{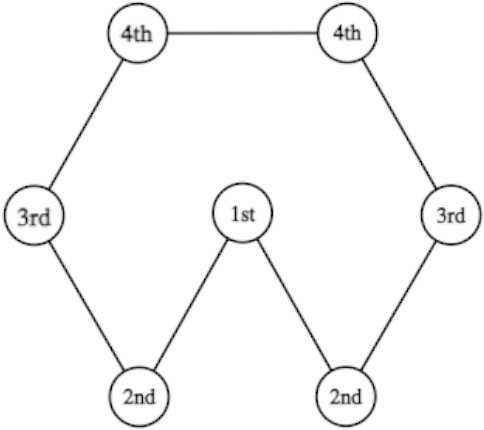}
\caption{\label{7cycle-simple-contagion} All nodes will be infected after 4 steps.}
\end{figure}

To illustrate the relationship between $L(C)$ and the length of $C$,  consider the following example. Suppose a simple contagion process in a cycle with seven nodes starts from one node and infects its neighbors with certainty ($q=1$). Then, in four infection steps, all nodes will be infected. Here, four is half of the cycle length rounded to the closest greater integer (Supplementary Figure\ref{7cycle-simple-contagion}).
In this scenario for an n-cycle, all nodes will be infected in $\lceil{n}/{2}\rceil$ steps, thus $L(C)$ equals half the length of the loop.

$L(C)$ in its simplest form is half the length of the cycle, which may vary with different contagion dynamics, providing an indicator that captures both the network structure and the contagion dynamics. However, the role of initially infected nodes in the cycles warrants more complex considerations, which we will demonstrate next.

When our network is not as simple as a single cycle or more than one node are initially infected, we need more modifications to accurately measure cycle length in relation to contagion dynamics. Consider the spread of the contagion from node A, already infected,  to node B in Supplementary Figure \ref{fig1_method}.
\begin{figure}[H]
\centering
\includegraphics[width=0.25\textwidth]{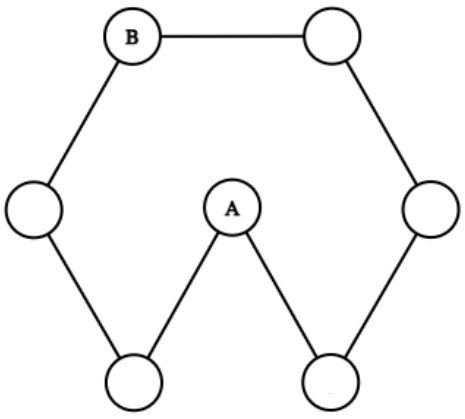}
\caption{\label{fig1_method}}
\end{figure}
In this scenario, if nodes C and D are also initially infected (Supplementary Figure \ref{fig2_method}), then they will influence the progression of contagion to node B. 
\begin{figure}[H]
\centering
\includegraphics[width=0.25\textwidth]{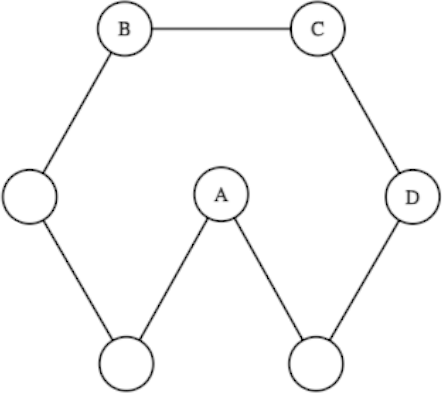}
\caption{\label{fig2_method}}
\end{figure}
Therefore, since nodes A, C, and D are all initially infected, we understand that the infection spreads not only from node A but also from the collective group of nodes A, C, and D. A potential solution to accurately model this dynamic is to consider these initially infected nodes as a single entity in our calculations.
\begin{figure}[H]
\centering
\includegraphics[width=0.25\textwidth]{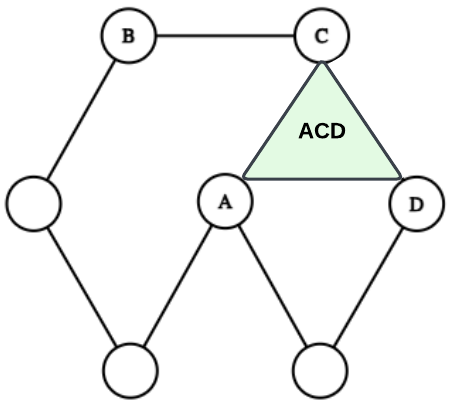}
\caption{\label{fig3_method}}
\end{figure}
Combining these nodes creates two smaller cycles, termed cycle 1 and cycle 2 in Supplementary Figure \ref{one_cycle_two_cycles}, that replace the original cycle. Since the spreading of the infection in these two cycles occurs independently and does not affect each other, we need to treat them separately. Our solution involves calculating $L(C)$ for both cycles and then taking the average over all such cycles.
\begin{figure}[H]
\centering
\includegraphics[width=0.25\textwidth]{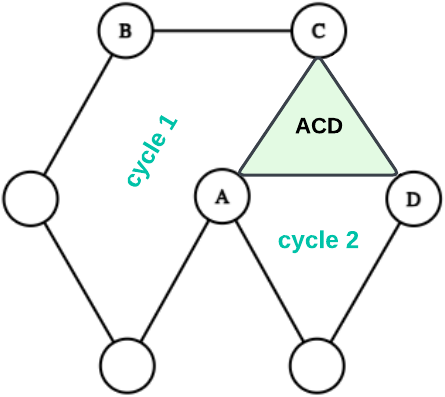}
\caption{\label{one_cycle_two_cycles}}
\end{figure}

Our construction of extended persistent homology (EPH) in SI section \ref{secSI:PHandEPH} is exactly similar to what we have done here. Especially because relative homology is the reduced homology of the quotient in our setting (see \cite{cohen2009extending} for more details).

\section{Persistent and Extended Persistent Homology}\label{secSI:PHandEPH}

\textbf{Persistent Homology.} Persistent homology is a framework for identifying and analyzing topological features in multiple dimensions, such as connected components (0-dimensional) and loops (1-dimensional). This approach utilizes a scalar function, known as filtration, to measure the prominence of these features within a graph structure. For a graph $G=(V,E)$ with the vertex set $V$ and the edge set $E$, we treat the set of all the vertices and edges, denoted by $X = V \bigcup E$, as a simplex. 

To construct a persistent homology, we set our filtration $f: X \to \mathbb{R}$  based on the time step at which the nodes get infected. For an edge $uv$, the filter function is calculated as $f(uv) = \max(f(u), f(v))$. The sublevel set $X_a$ consists of all simplices in $X$ for which the filter function does not exceed $a$: $X_a = \{x \in X | f(x) \leq a\}$. As $a$ varies from $-\infty$ to $\infty$, it generates an ascending filtration sequence from $X_{-\infty}$ to $X_{\infty}$, revealing a spectrum of topological features as they appear and vanish.

The emergence and integration of these features is meticulously tracked throughout the filtration. For example, as shown in Supplementary Figure \ref{example_PH}, two topological features, a connected component (dim 0) and a loop (dim 1), emerge at times 1 and 3, respectively, and do not vanish, leading us to consider their death time as \(\infty\).
\begin{figure}[H]
\centering
\includegraphics[width=0.8\textwidth]{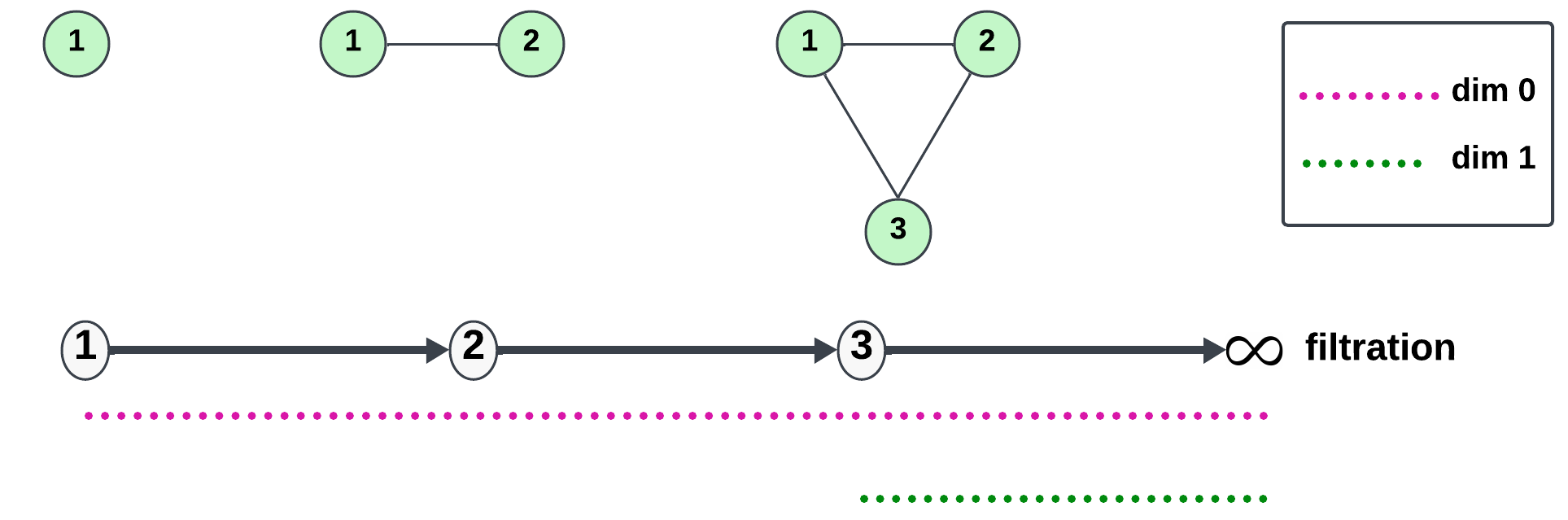}
\caption{\label{example_PH} Example of persistent homology and filtration. The plot illustrates the topological features of dimensions zero (connected components) and one (loops). Such plots are referred to as \textit{barcodes} in topological data analysis (TDA) literature.}
\end{figure}
The analysis employs the homology functor on these filtration to construct a persistence diagram (PD), a two-dimensional scatter plot that records the birth and death times ($b$ and $d$, respectively) of topological features. The \textit{lifetime} or \textit{persistence} $|d - b|$ of these features, which indicates their relative importance and stability during filtration, are thus plotted. 
\begin{figure}[H]
\centering
\includegraphics[width=0.5\textwidth]{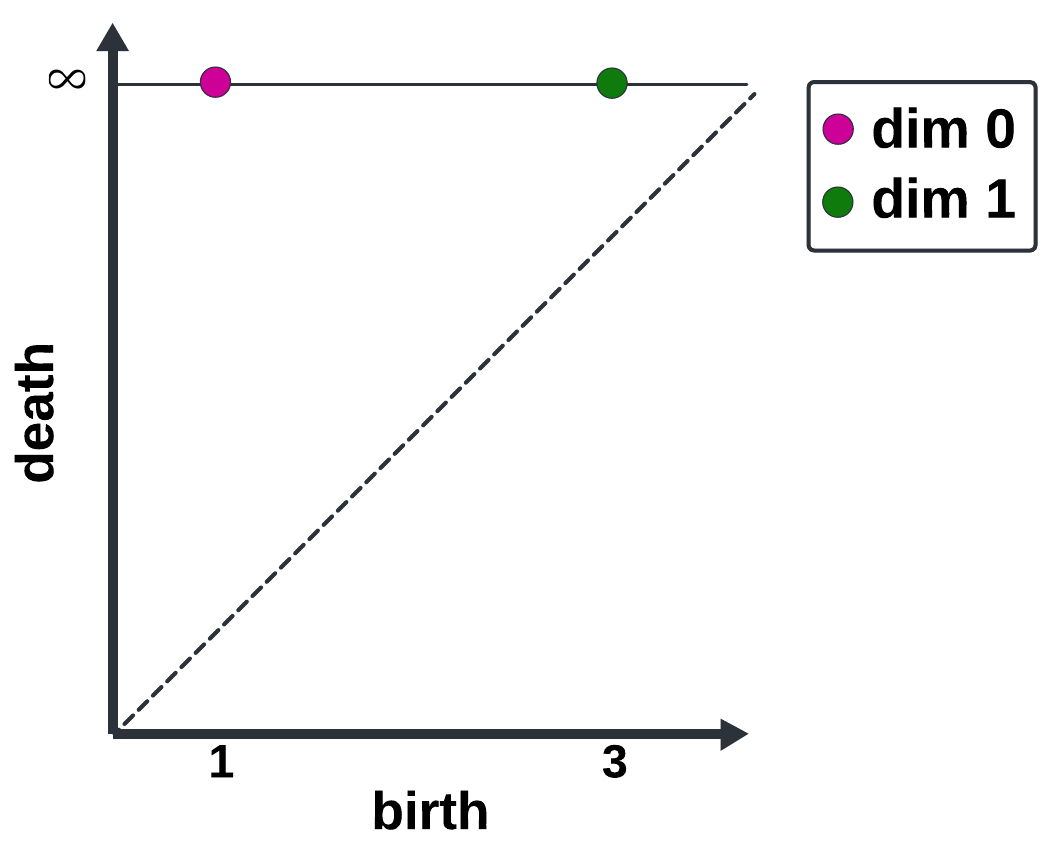}
\caption{\label{example_PD} Example of a persistent diagram. This is the persistent diagram corresponding to the shape shown in Supplementary Figure \ref{example_PH}. The diagram illustrates when topological features in different dimensions -- dimension zero for connected components and dimension one for loops -- appear and disappear. This method of representation is known as a \textit{persistent diagram} in the topological data analysis literature.
}
\end{figure}
This diagram serves as a concise and effective tool for representing the persistence and significance of topological structures relative to the initial filter function applied to the graph.

\textbf{Extended Persistent Homology.} In standard persistent homology, each feature of the domain (such as a graph) is observed to have a point of origin or ``birth". These features may persist indefinitely, with no ``death" time (symbolically, the death time is infinite). For example, you can consider the only loop that is generated at $3$ and never vanishes in Supplementary Figure \ref{example_PD}. These everlasting features are termed essential features. Particularly in graph theory, one-dimensional essential features, which correspond to independent loops, are not effectively captured in ordinary persistence methods.

To address this, an extended persistence module is introduced, which enhances our ability to gauge the significance of these one-dimensional essential features. This is represented as a sequence of homology groups over different thresholds:
$$
\begin{aligned}
& \emptyset=H\left(X_{-\infty}\right) \rightarrow \cdots \rightarrow H\left(X_a\right) \rightarrow \cdots \\ &\rightarrow H(X)=H\left(X, X^{\infty}\right) \rightarrow  \cdots \rightarrow H\left(X, X^a\right) \rightarrow \\
& \cdots \rightarrow H\left(X, X^{-\infty}\right),
\end{aligned}
$$
where $X^a = \{ x \in X \mid f(x) \geq a \}$ defines a superlevel set at value $a$ and $H(X, X^{a})$ means relative homology group of the pair of spaces $(X, X^{a})$. Intuitively, you can think of $H(X, X^{a})$ as a homology of $X/X^a$ where all elements of $X^a$ are identified with each other. The second part of this sequence, which involves a descending filtration from $H(X, X^\infty)$ to $H(X, X^{-\infty})$, captures the features as they diminish and eventually die out. The trivial end homology $H(X, X^{-\infty})$, which indicates that all loop features eventually dissolve, is also captured by the persistence diagram endpoint. Extended persistence homology improves the analysis by showing the complete life cycle of each feature \cite{yan2022neural}. The following figure shows what happens in the extended persistent homology for the example in Supplementary Figure \ref{example_PH}:
\begin{figure}[H]
\centering
\includegraphics[width=0.8\textwidth]{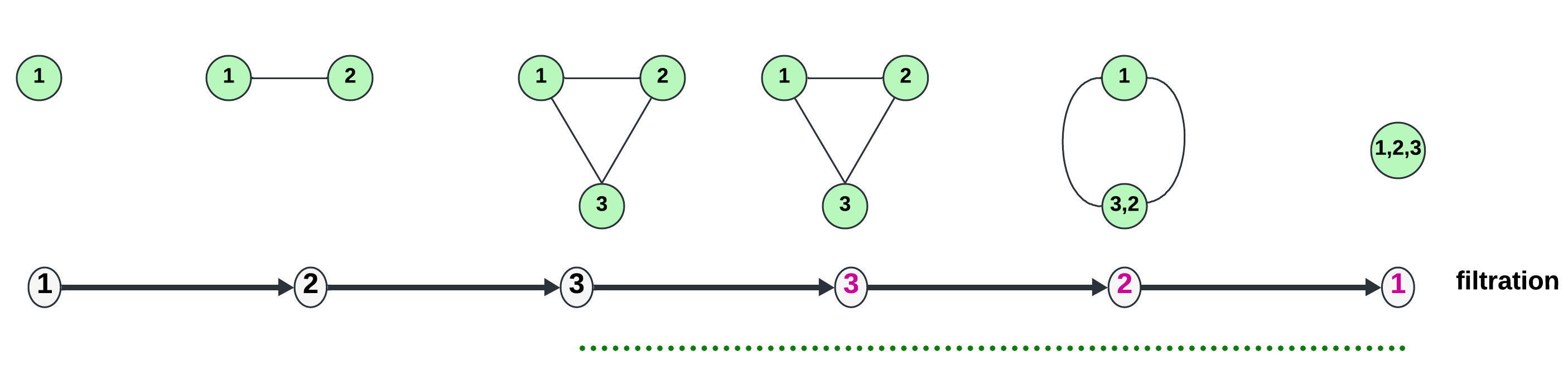}
\caption{\label{example_EPH} Example of extended persistent homology. The plot illustrates the extended persistence homology of dimension one (loops).}
\end{figure}

In the case of extended persistent homology, when only dimension one is considered (using the $H_1$ functor), the lifetime of the generators deals specifically with the lifetime of loops or cycles. As mentioned above, intuitively we need to define lifetime as the measurement of the time required for an infection to spread around a cycle. Therefore, we use the time step at which the nodes become infected as our filtration. In this setup, when employing extended persistent homology, we first merge all elements of $X^a$ together to find $H(X, X^{a})$ and then calculate the lifetime of the cycles. 

\section{Distribution of Topological Features for Other Datasets}\label{secSI:distributions}
In this section, we present the EPH distributions for various datasets, analogous to the results for the email dataset shown in Figure \ref{fig:distributions}. In addition, we provide a sample where the value of $q$ for complex contagion increases to $0.04$, to demonstrate that the differences between the EPH distributions for simple and complex contagion remain significant. All simulation parameters are the same as in Figure \ref{fig:distributions} of the main paper. Here, we present the distributions for the conference dataset:
\begin{figure}[H]
\centering
\includegraphics[width=1\textwidth]{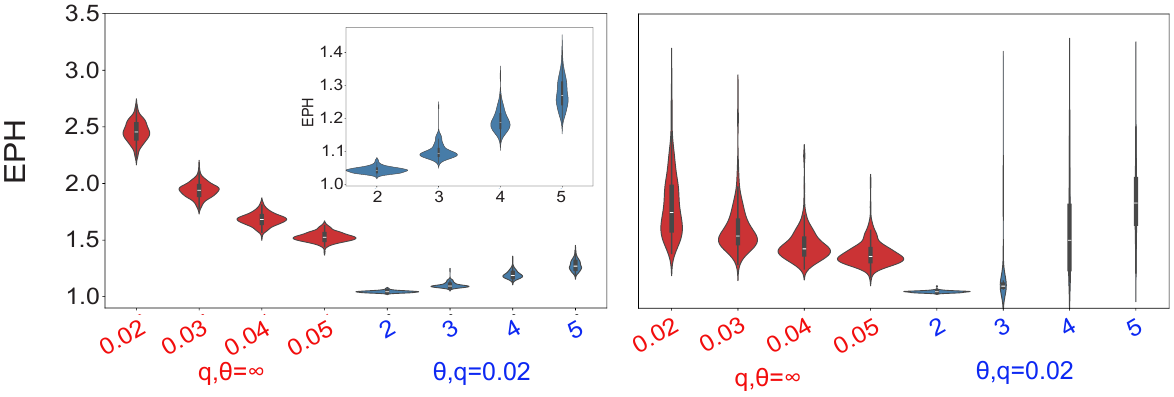}
\caption{\label{dist_conf} EPH distribution for the conference network}
\end{figure}
And same plot for school dataset:
\begin{figure}[H]
\centering
\includegraphics[width=1\textwidth]{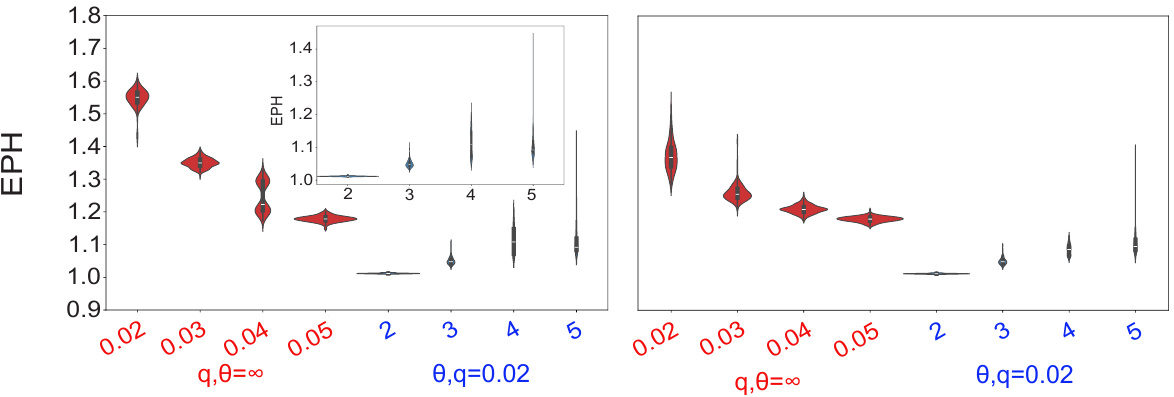}
\caption{\label{dist_school} EPH distribution for the school network}
\end{figure}

\section{Effect of Increasing \texorpdfstring{$q$}{q} on EPH in Complex Contagions}
Here, we demonstrate the effect of increasing the value of $q$ in complex contagion to $0.04$ to test whether our classifier and regression algorithms function effectively at higher $q$ values. As evident in the distributions below, the EPH differs significantly between simple and complex contagion, even at higher $q$ values. All other parameters are identical to those used in Figure \ref{fig:distributions}.
\begin{figure}[H]
\centering
\includegraphics[width=1\textwidth]{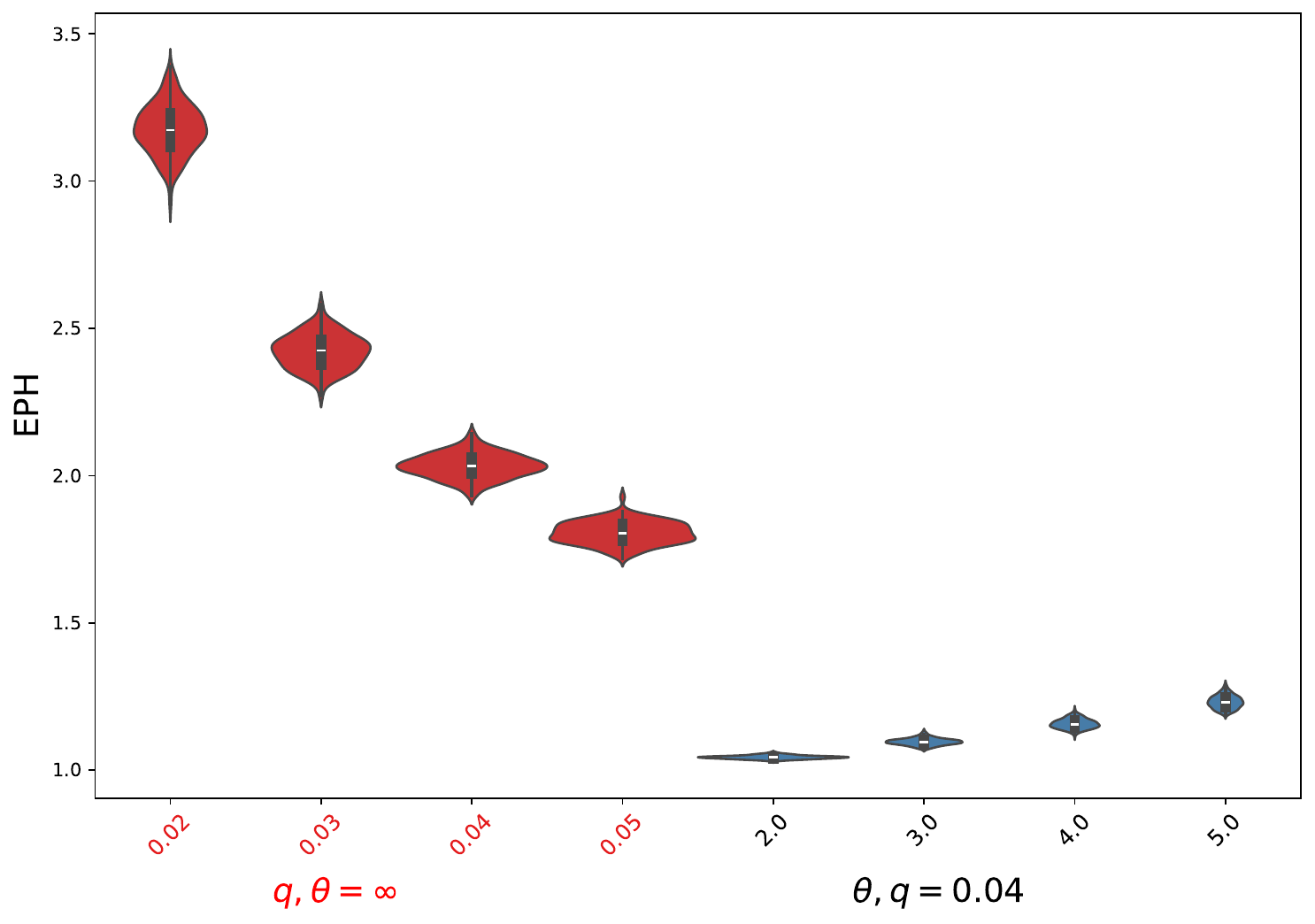}
\caption{\label{dist_email_q_4} EPH distribution for the email network with $q=0.04$ in complex contagion}
\end{figure}
\begin{figure}[H]
\centering
\includegraphics[width=1\textwidth]{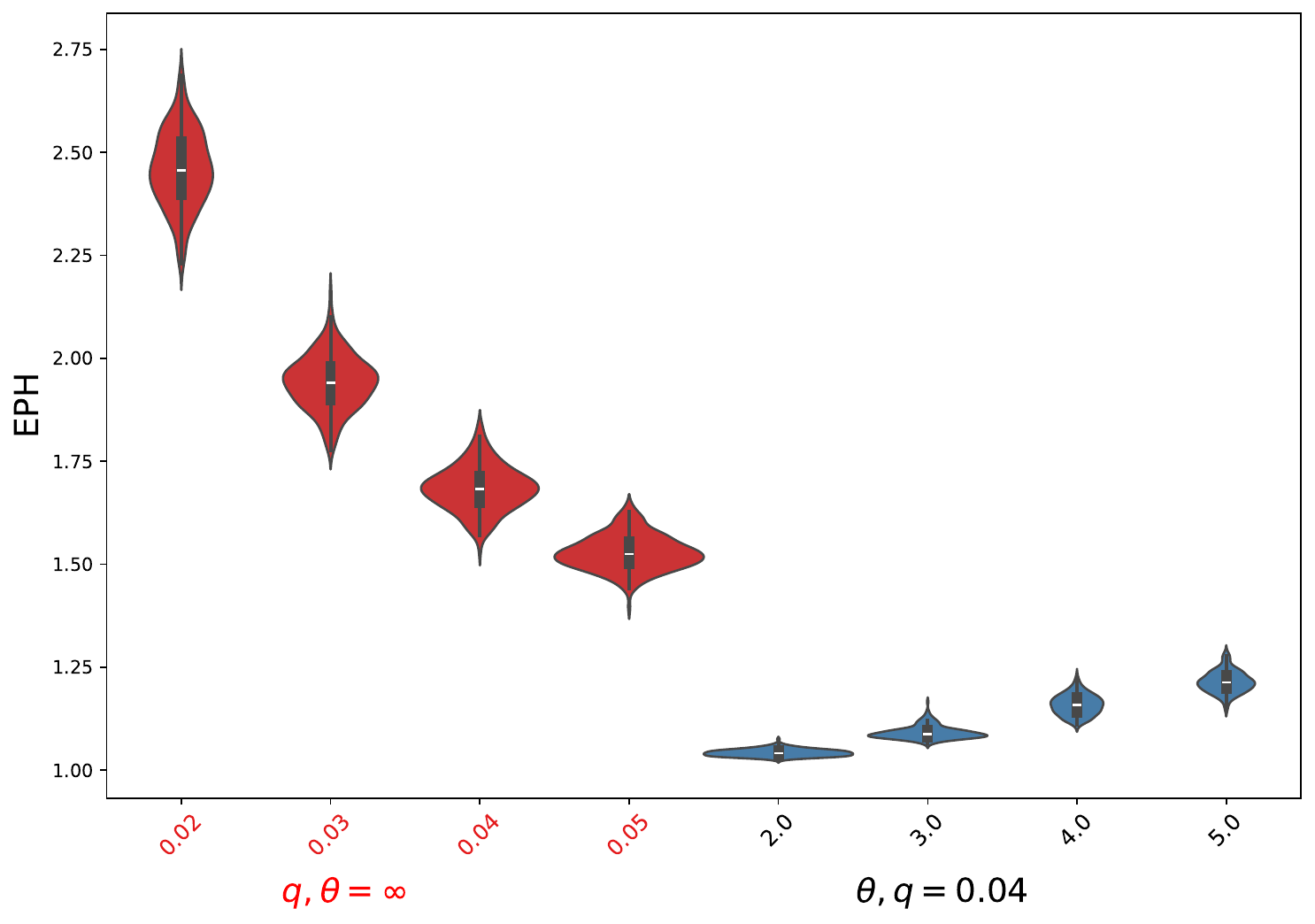}
\caption{\label{dist_conf_q_4} EPH distribution for the conference network with $q=0.04$ in complex contagion}
\end{figure}
\begin{figure}[H]
\centering
\includegraphics[width=1\textwidth]{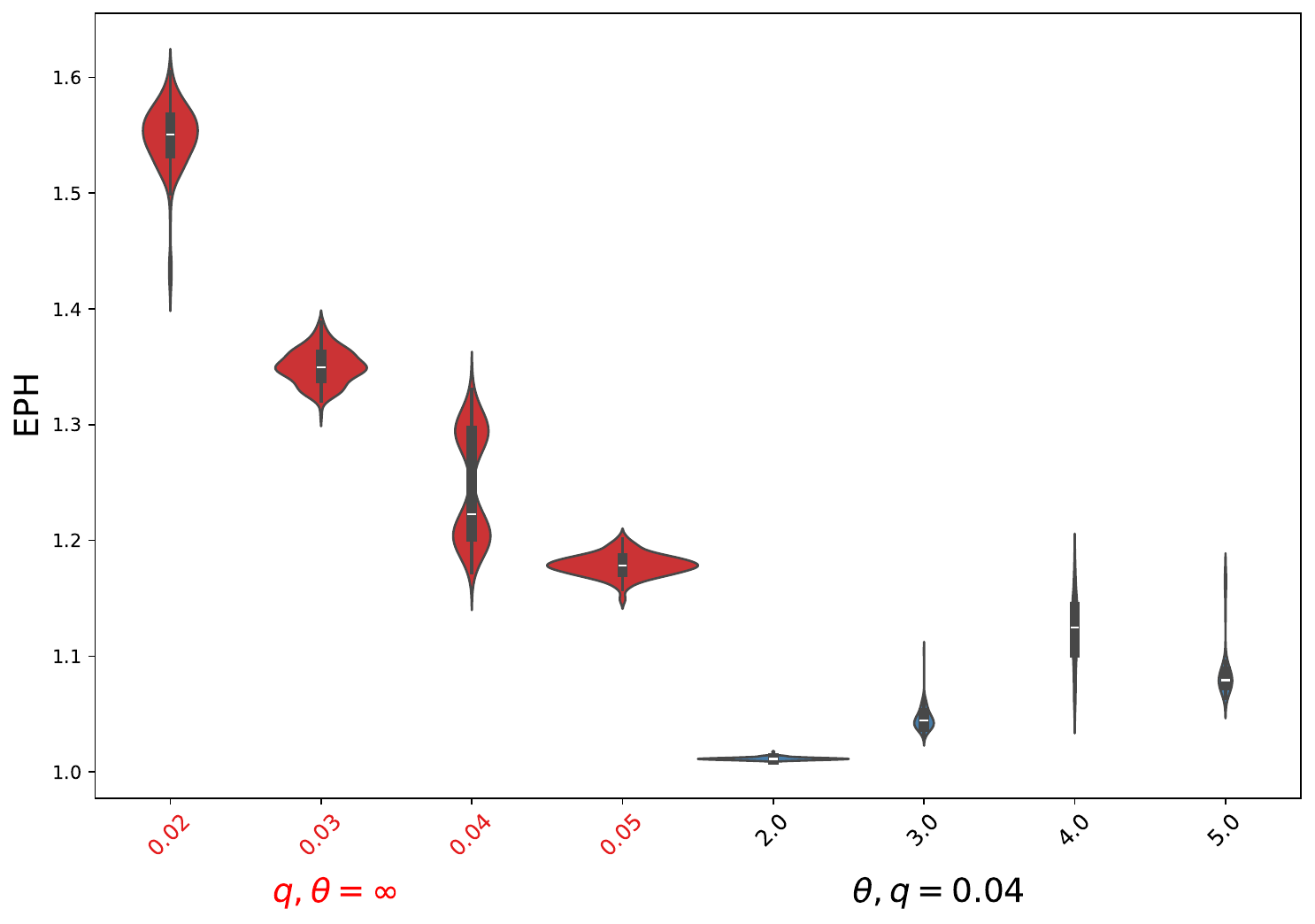}
\caption{\label{dist_school_q_4} EPH distribution for the school network with $q=0.04$ in complex contagion}
\end{figure}

\section{Classification and Regression Results for Other Datasets}\label{SIsec:other-results}
In this section, we present the classification and regression results for additional datasets. We include the classification between simple and complex contagion, as well as the regression analysis to predict $\theta$ in complex contagion and $q$ in simple contagion. These are analogous to the results shown in Figure \ref{fig:results} of the main text, but are applied to two other datasets: conference and school. We begin with the classification results followed by the regression analysis.

\begin{figure}[H]
\centering
\includegraphics[width=1\textwidth]{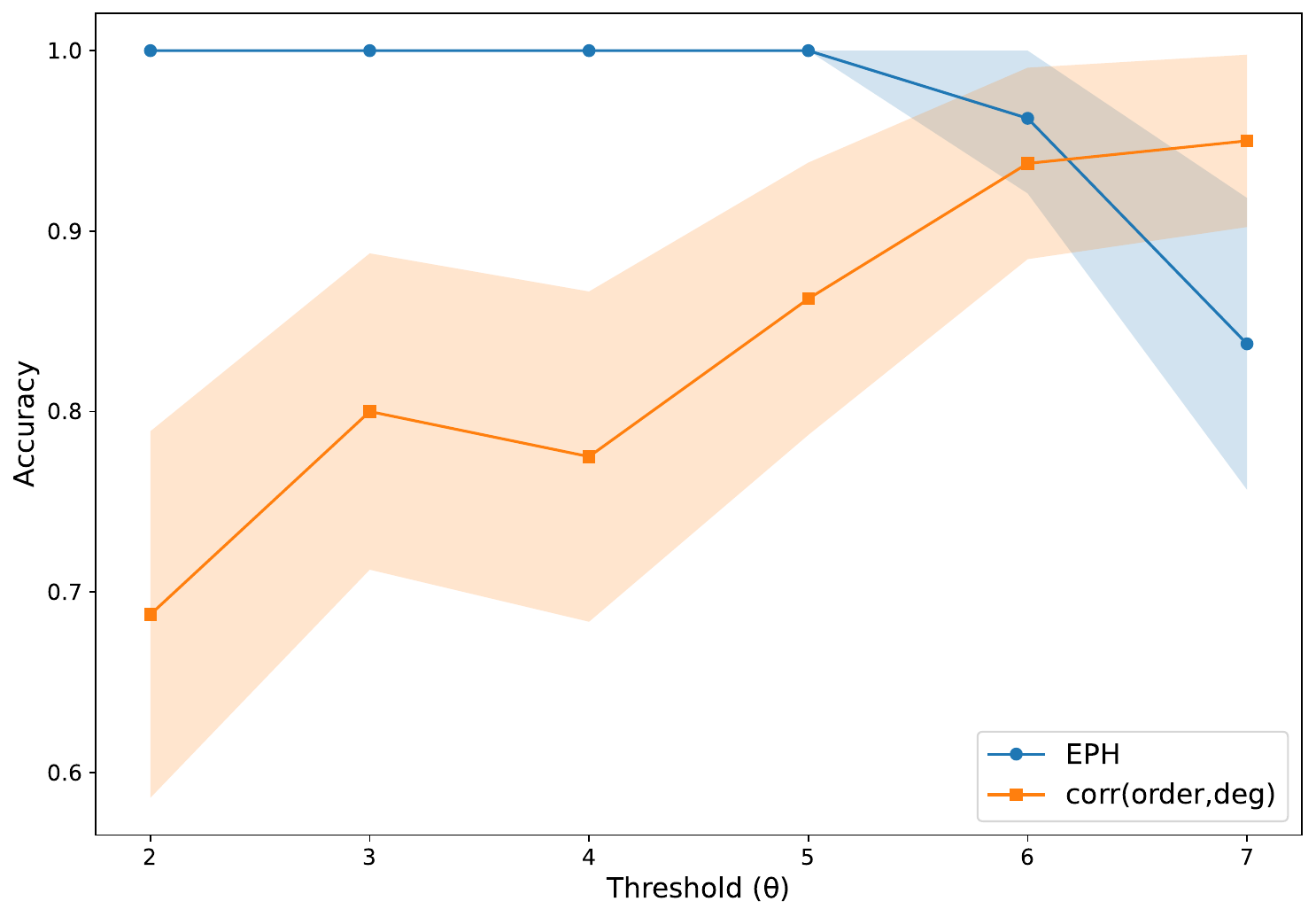}
\caption{\label{class_conf} Classification results for the conference dataset at different $\theta$ values.}
\end{figure}

\begin{figure}[H]
\centering
\includegraphics[width=1\textwidth]{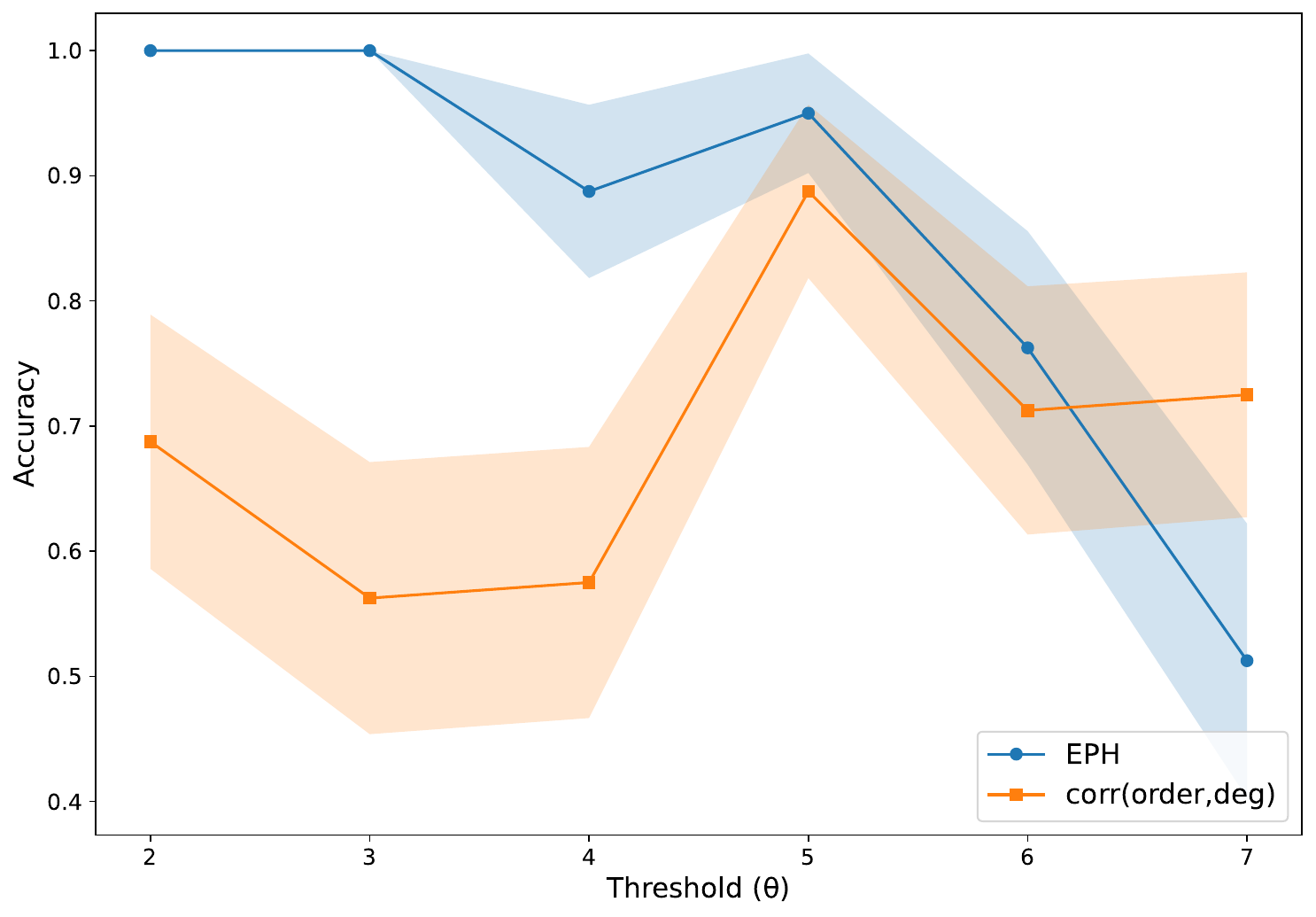}
\caption{\label{class_school} Classification results for the school dataset at different $\theta$ values.}
\end{figure}

As is apparent, our classification performance is excellent in all settings except for high values of $\theta$. In this scenario, most infections occur due to the simple contagion mechanism, as $q$ is set to 0.02 for complex contagion, which complicates the classification task.

In addition, with increasing $q$ for complex contagion, classification performance remains perfect. The following figure shows the same results with $q$ increased to $0.04$.

\begin{figure}[H]
\centering
\includegraphics[width=1\textwidth]{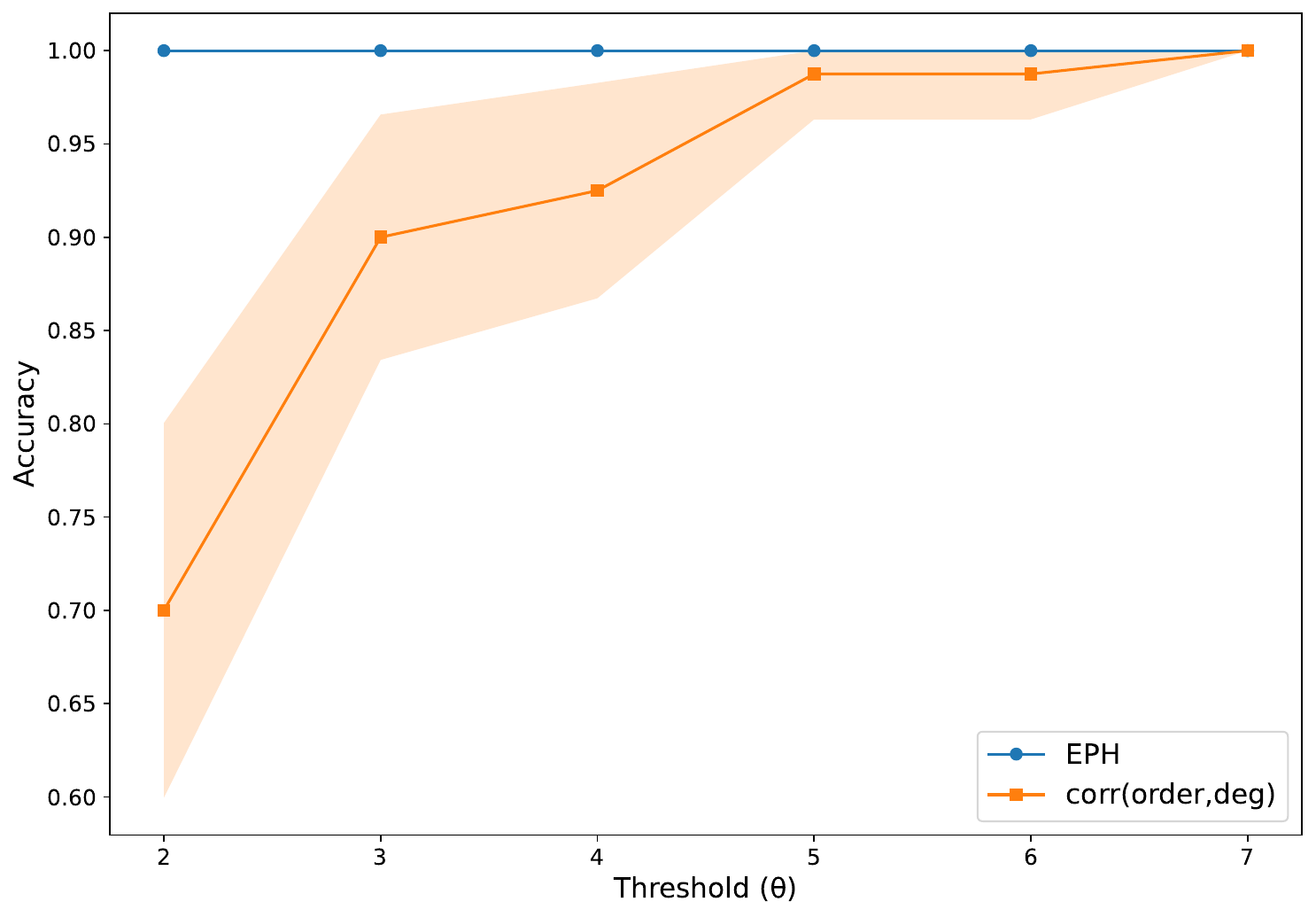}
\caption{\label{class_email_q} Classification results for the email dataset at different $\theta$ values and with $q = 0.04$ in complex contagion.}
\end{figure}

\begin{figure}[H]
\centering
\includegraphics[width=1\textwidth]{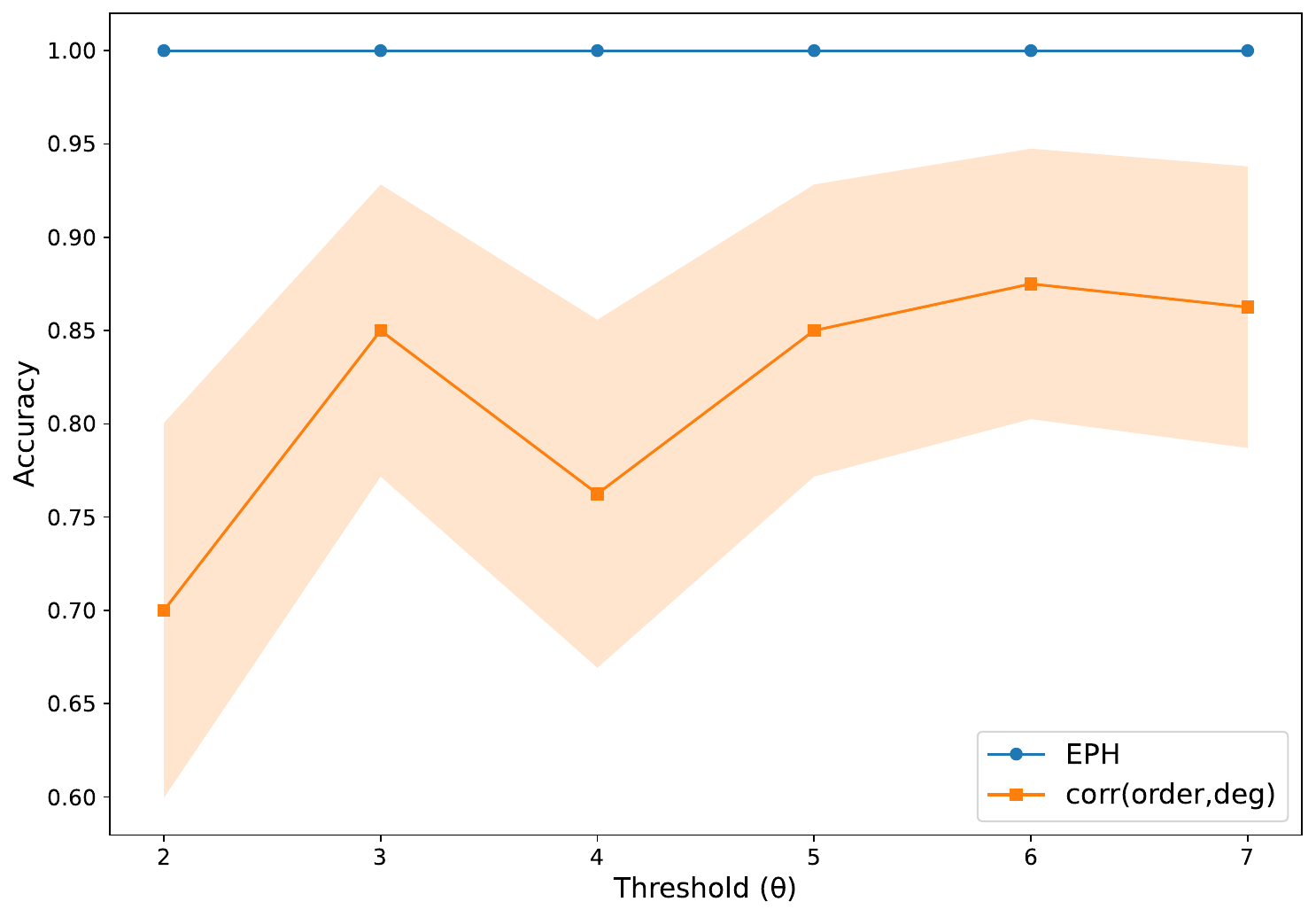}
\caption{\label{class_conf_q} Classification results for the conference dataset at different $\theta$ values and with $q = 0.04$ in complex contagion.}
\end{figure}

\begin{figure}[H]
\centering
\includegraphics[width=1\textwidth]{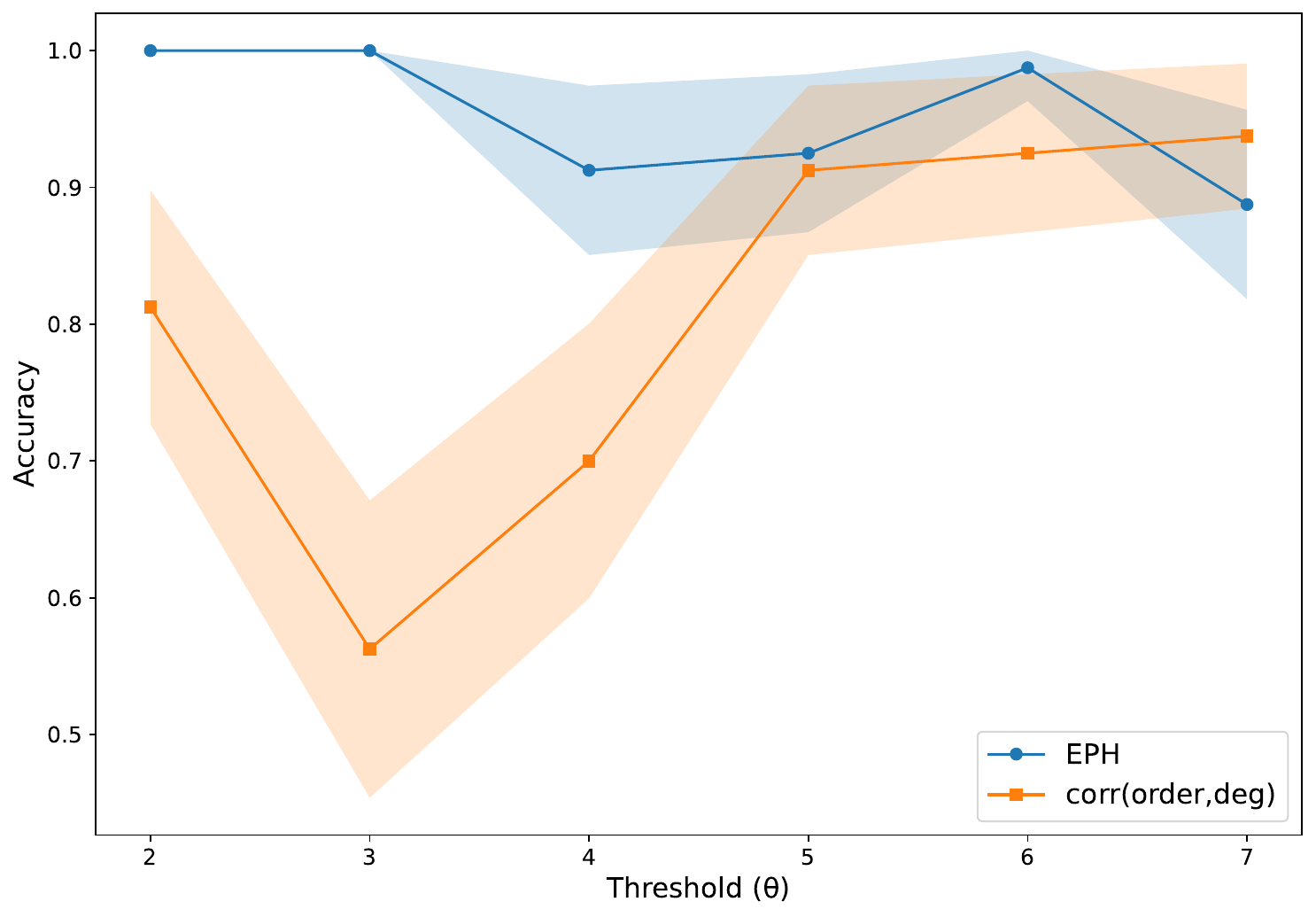}
\caption{\label{class_school_q} Classification results for the school dataset at different $\theta$ values and with $q = 0.04$ in complex contagion.}
\end{figure}

Here are the results for the $\theta$ regression analysis:
\begin{figure}[H]
\centering
\includegraphics[width=1\textwidth]{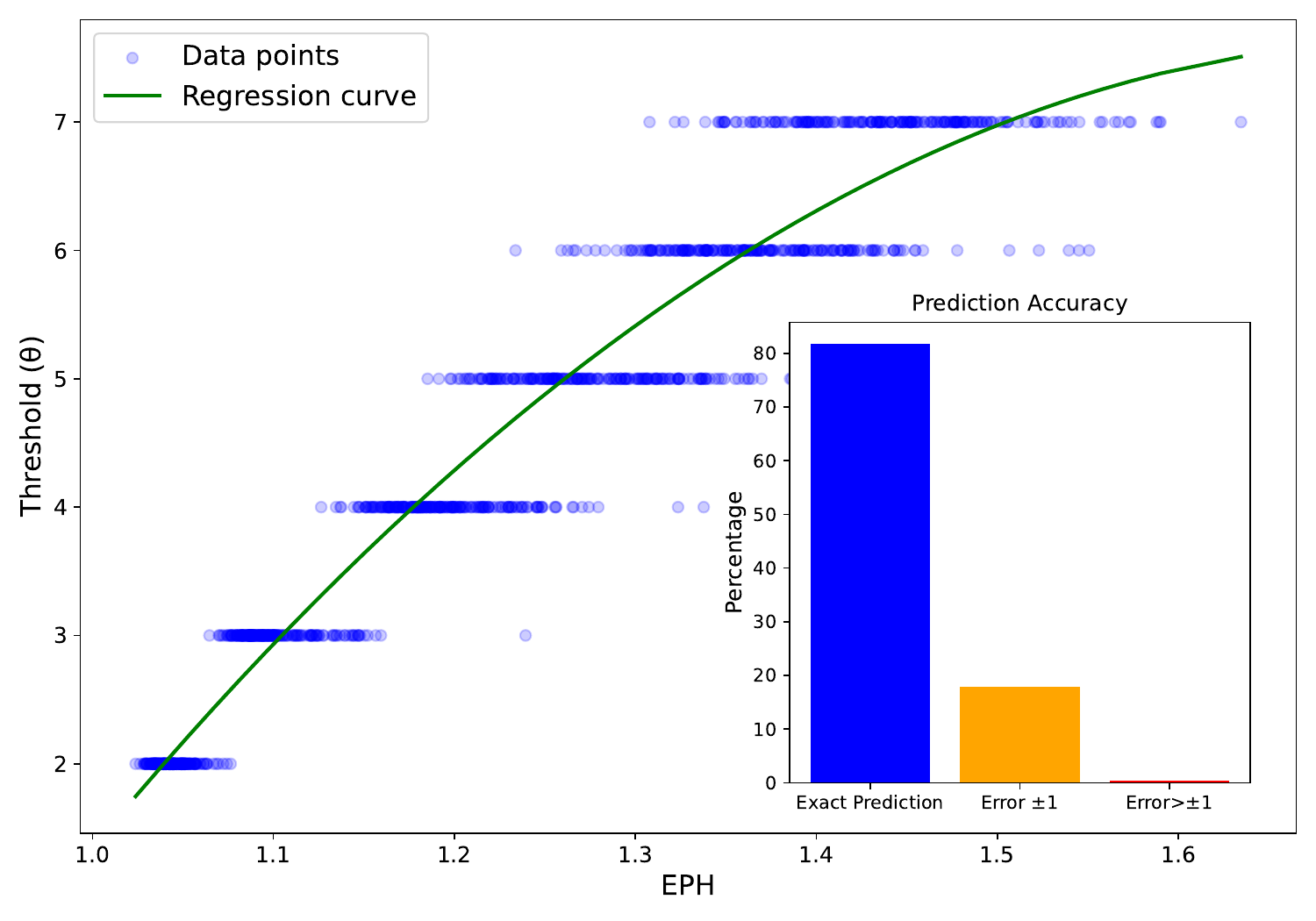}
\caption{\label{tr_conf} Complex contagion regression results for the conference dataset.}
\end{figure}

\begin{figure}[H]
\centering
\includegraphics[width=1\textwidth]{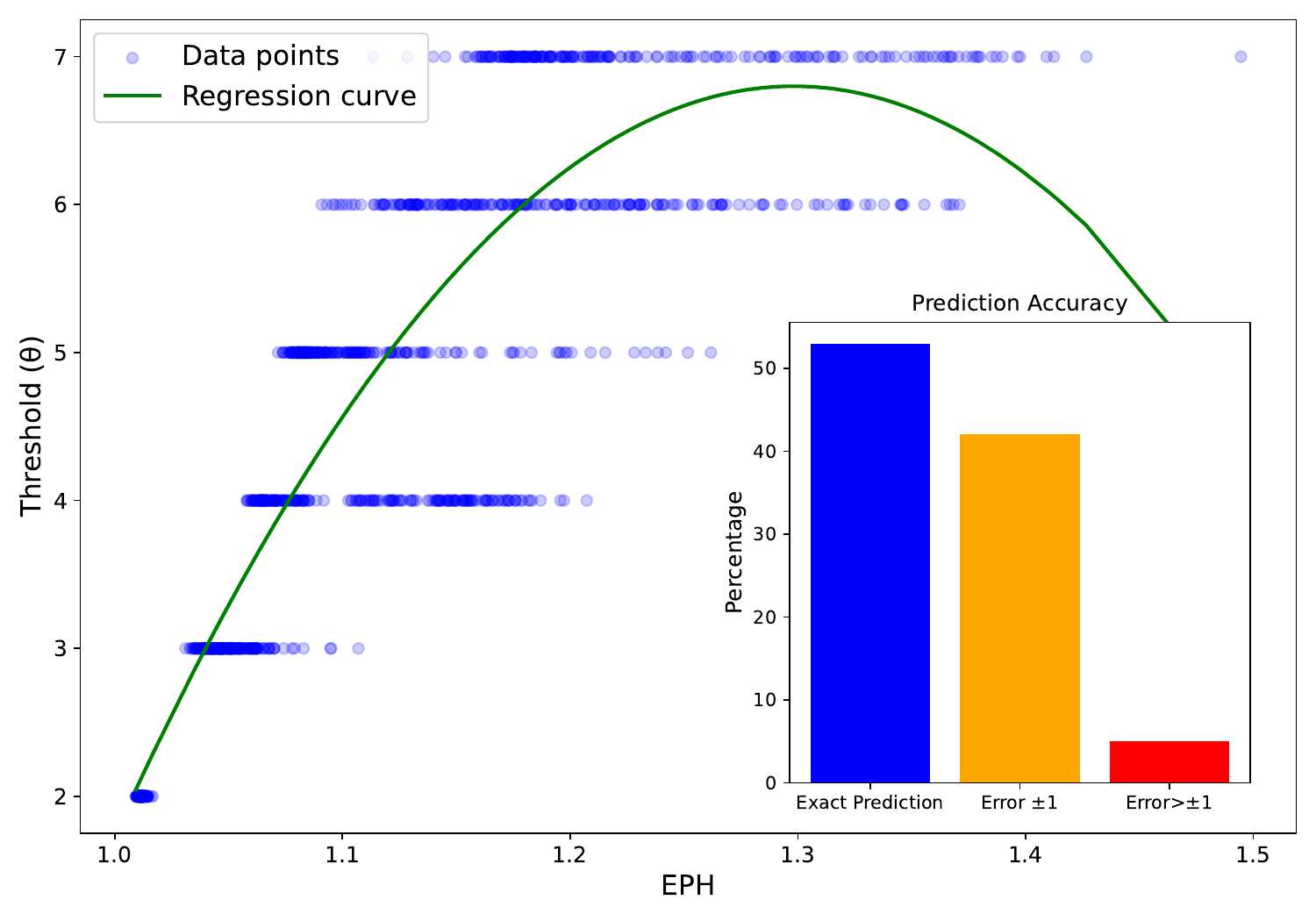}
\caption{\label{tr_school} Complex contagion regression results for the school dataset.}
\end{figure}

And here is the regression of $q$ for simple contagion:
\begin{figure}[H]
\centering
\includegraphics[width=1\textwidth]{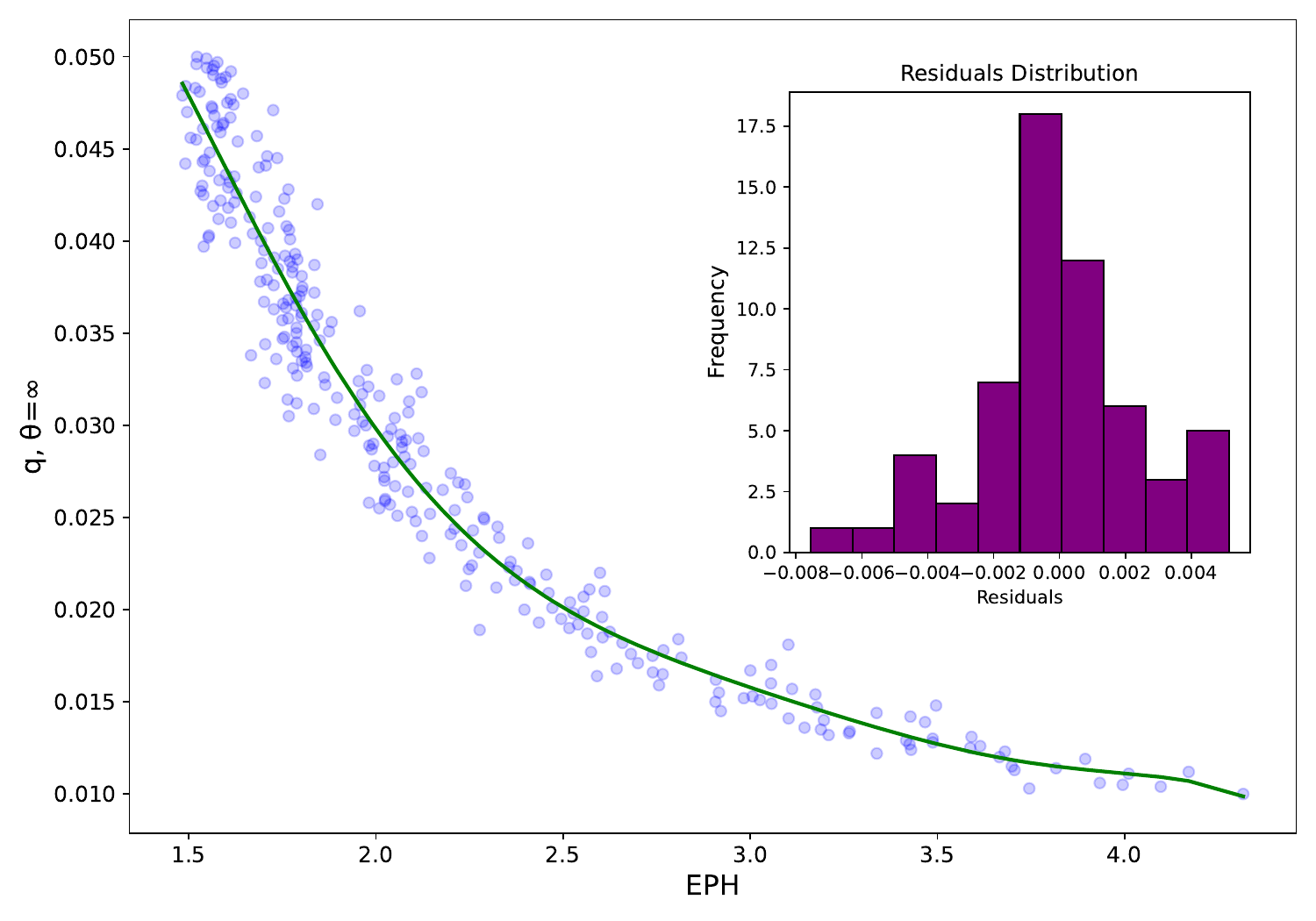}
\caption{\label{tq_conf} Simple contagion regression results for the conference dataset.}
\end{figure}

\begin{figure}[H]
\centering
\includegraphics[width=1\textwidth]{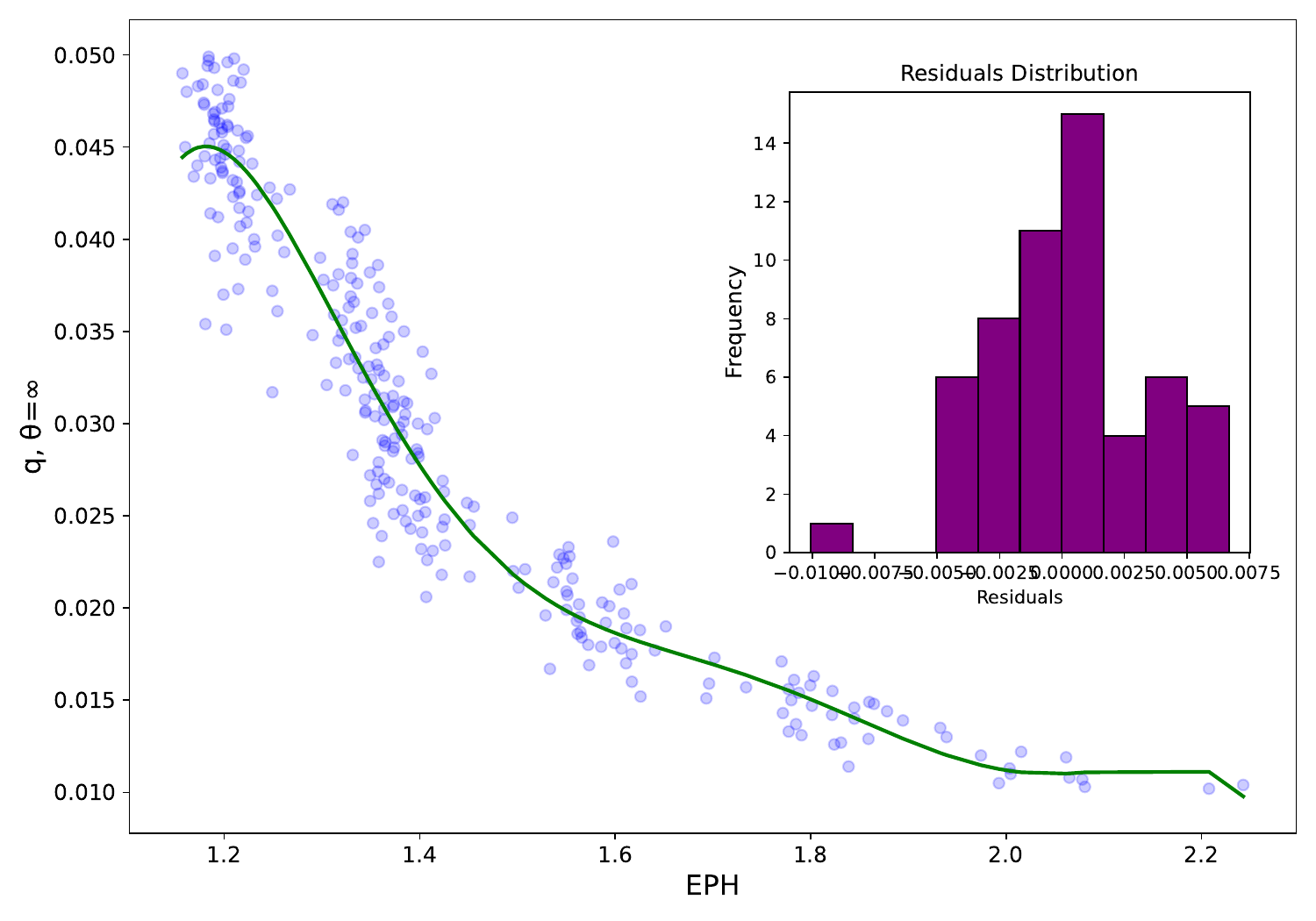}
\caption{\label{tq_school} Simple contagion regression results for the school dataset.}
\end{figure}

\section{Numerical Implementation Details}\label{secSI:details_implementation}
In this section, we explain the numerical and computational procedures used to generate the results of this paper.

\subsection{Computing extended persistent homology}\label{secSI:details_implementation:eph}
For calculating extended persistent homology, we utilized the software package described in \cite{gudhi:urm}. This package employs a Python function to compute extended persistent homology. To implement extended persistence, the package uses the coning technique, which is shown to be roughly equivalent to relative homology, as proven in \cite{cohen2009extending}. The algorithm is as follows:

\begin{algorithm}[H]
\caption{Computing Extended Persistent Homology}
\begin{algorithmic}[1]
\State Map all ordinary filtration values within the interval \([-2, -1]\).
\State Cone the entire simplicial complex.
\State Map all coned filtration values to \([1, 2]\) and assign the vertex used for coning a filtration value of \(-3\).
\State Proceed with calculations as in ordinary persistent homology.
\end{algorithmic}
\end{algorithm}

\subsection{Variance reduction techniques in contagion simulations}
 For each scenario, we performed 500 simulations with uniformly chosen parameters, except for the parameter being studied, which was kept constant. For example, in the simulations for Panel A of Figure \ref{fig:results}, we ran 500 simulations for each value of $q$, with other parameters chosen uniformly at random but consistently throughout all runs to minimize variance. Initially, infected nodes were uniformly selected from the set of all ten-element subsets of nodes, ensuring that each potential subset of nodes had an equal probability of being chosen. Parameters such as $\beta$ and $\theta$ were also uniformly selected from predefined sets. For simulations involving limited information, such as those with a restricted number of infection steps, the procedure was similar, except that the infection steps were capped at the values indicated on the x-axis of Panel A in Figure \ref{fig_liminf:results}. In scenarios with limited information, while contagion steps were performed on the entire graph, the calculations of the EPH features were based on the observed parts of the graph, whose size is indicated on the x axis of panel B in Figure \ref{fig_liminf:results}.

\end{document}